\documentclass{mybookalone}
\usepackage{makeidx,epsfig}
\usepackage{amsthm,amsmath,amssymb}
\usepackage{setspace,graphicx}
\usepackage{Generic}
\usepackage[sort,longnamesfirst]{natbib}

\def\beq{\begin{eqnarray}}
\def\eeq{\end{eqnarray}}
\def\T{^{\scriptscriptstyle T}}
\def\E{\mathbb{E}\,}

\numberwithin{equation}{section}
\theoremstyle{plain}

\begin{document}


\chapter[MCMC using Hamiltonian dynamics]{MCMC using Hamiltonian dynamics}

\begin{Large}\vspace*{-0.3in}
{\em Radford M.\ Neal, University of Toronto}
\end{Large}

\vspace{12pt}

\begin{quotation} 

\noindent Hamiltonian dynamics can be used to produce distant
proposals for the Metropolis algorithm, thereby avoiding the slow
exploration of the state space that results from the diffusive
behaviour of simple random-walk proposals.  Though originating in
physics, Hamiltonian dynamics can be applied to most problems with
continuous state spaces by simply introducing fictitious ``momentum''
variables.  A key to its usefulness is that Hamiltonian dynamics
preserves volume, and its trajectories can thus be used to define
complex mappings without the need to account for a hard-to-compute
Jacobian factor --- a property that can be exactly maintained even
when the dynamics is approximated by discretizing time.  In this
review, I discuss theoretical and practical aspects of Hamiltonian
Monte Carlo, and present some of its variations, including using
windows of states for deciding on acceptance or rejection, computing
trajectories using fast approximations, tempering during the course of
a trajectory to handle isolated modes, and short-cut methods that
prevent useless trajectories from taking much computation time.

\end{quotation}

\vspace{-3pt}

\section{Introduction}\label{s:intro}

Markov Chain Monte Carlo (MCMC) originated with the classic paper of
\citet{metropolis:1953}, where it was used to simulate the
distribution of states for a system of idealized molecules.  Not long
after, another approach to molecular simulation was introduced
\citep{alder:1959}, in which the motion of the molecules was
deterministic, following Newton's laws of motion, which have an
elegant formalization as \textit{Hamiltonian dynamics}.  For finding
the properties of bulk materials, these approaches are asymptotically
equivalent, since even in a deterministic simulation, each local
region of the material experiences effectively random influences from
distant regions.  Despite the large overlap in their application
areas, the MCMC and molecular dynamics approaches have continued to
co-exist in the following decades \citep[see][]{frenkel:1996}.

In \citeyear{duane:1987}, a landmark paper by Duane, Kennedy,
Pendleton, and Roweth united the MCMC and molecular dynamics
approaches.  They called their method ``Hybrid Monte Carlo'', which
abbreviates to ``HMC'', but the phrase ``Hamiltonian Monte Carlo'',
retaining the abbreviation, is more specific and descriptive, and I
will use it here.  Duane, \textit{et al.} applied HMC not to molecular
simulation, but to lattice field theory simulations of quantum
chromodynamics.  Statistical applications of HMC began with my use of
it for neural network models \citep{neal:1996}.  I also provided a
statistically-oriented tutorial on HMC in a review of MCMC methods
\citep[Chapter 5]{neal:1993}.  There have been other applications of
HMC to statistical problems \citep[eg,][]{ishwaran:1999,schmidt:2009} and
statistically-oriented reviews \citep[eg,][Chapter~9]{Liu:2001}, but
HMC still seems to be under-appreciated by statisticians, and perhaps
also by physicists outside the lattice field theory community.

This review begins by describing Hamiltonian dynamics.  Despite
terminology that may be unfamiliar to non-physicists, the features of
Hamiltonian dynamics that are needed for HMC are elementary.  The
differential equations of Hamiltonian dynamics must be discretized for
computer implementation.  The ``leapfrog'' scheme that is typically
used is quite simple.

Following this introduction to Hamiltonian dynamics, I describe how to
use it to construct a Markov chain Monte Carlo method.  The first step
is to define a Hamiltonian function in terms of the probability
distribution we wish to sample from.  In addition to the variables we
are interested in (the ``position'' variables), we must introduce
auxiliary ``momentum'' variables, which typically have independent
Gaussian distributions.  The Hamiltonian Monte Carlo method alternates
simple updates for these momentum variables with Metropolis updates in
which a new state is proposed by computing a trajectory according to
Hamiltonian dynamics, implemented with the leapfrog method.  A state
proposed in this way can be distant from the current state but
nevertheless have a high probability of acceptance.  This bypasses the
slow exploration of the state space that occurs when Metropolis
updates are done using a simple random-walk proposal distribution.
(An alternative way of avoiding random walks is to use short
trajectories but only partially replace the momentum variables between
trajectories, so that successive trajectories tend to move in the same
direction.)

After presenting the basic HMC method, I discuss practical issues of
tuning the leapfrog stepsize and number of leapfrog steps, as well as
theoretical results on the scaling of HMC with dimensionality.  I then
present a number of variations on HMC.  The acceptance rate for HMC
can be increased for many problems by looking at ``windows'' of states
at the beginning and end of the trajectory.  For many statistical
problems, approximate computation of trajectories (eg, using subsets
of the data) may be beneficial.  Tuning of HMC can be made easier
using a ``short-cut'' in which trajectories computed with a bad choice
of stepsize take little computation time.  Finally, ``tempering''
methods may be useful when multiple isolated modes exist.

\section{Hamiltonian dynamics}\label{s:ham}

Hamiltonian dynamics has a physical interpretation that can provide
useful intuitions.  In two dimensions, we can visualize the dynamics
as that of a frictionless puck that slides over a surface of varying
height.  The state of this system consists of the \textit{position} of
the puck, given by a 2D vector $q$, and the \textit{momentum} of the
puck (its mass times its velocity), given by a 2D vector $p$.  The
\textit{potential energy}, $U(q)$, of the puck is proportional to the
height of the surface at its current position, and its \textit{kinetic
energy}, $K(p)$, is equal to $|p|^2/(2m)$, where $m$ is the mass of
the puck.  On a level part of the surface, the puck moves at a
constant velocity, equal to $p/m$.  If it encounters a rising slope,
the puck's momentum allows it to continue, with its kinetic energy
decreasing and its potential energy increasing, until the kinetic
energy (and hence $p$) is zero, at which point it will slide back down
(with kinetic energy increasing and potential energy decreasing).

In non-physical MCMC applications of Hamiltonian dynamics, the
position will correspond to the variables of interest.  The potential
energy will be minus the log of the probability density for these
variables.  Momentum variables, one for each position variable, will
be introduced artificially.

These interpretations may help motivate the exposition below, but if
you find otherwise, the dynamics can also be understood as simply
resulting from a certain set of differential equations.

\subsection{Hamilton's equations}\label{ss:eq}

Hamiltonian dynamics operates on a $d$-dimensional \textit{position}
vector, $q$, and a $d$-dimensional \textit{momentum} vector, $p$, so
that the full state space has $2d$ dimensions.  The system is described
by a function of $q$ and $p$ known as the \textit{Hamiltonian}, $H(q,p)$.

\paragraph{Equations of motion.}
The partial derivatives of the Hamiltonian determine how $q$ and $p$ change 
over time, $t$, according to Hamilton's equations:
\beq
  {dq_i \over dt} & = & {\partial H \over \partial p_i}
    \label{e:ham-eq-a1}\\[4pt]
  {dp_i \over dt} & = & - {\partial H \over \partial q_i} 
    \label{e:ham-eq-a2}
\eeq
for $i = 1,\ldots,d$.  For any time interval of duration $s$, these equations
define a mapping, $T_s$, from the state at any time $t$ to the state 
at time $t+s$.  (Here, $H$, and hence $T_s$, are assumed to not depend on 
$t$.)  

Alternatively, we can combine the vectors $q$ and $p$ into the vector
$z = (q, p)$ with $2d$ dimensions, and write Hamilton's equations as
\beq
  {dz \over dt} & = & J\, \nabla H(z)
    \label{e:ham-eqz}
\eeq
where $\nabla H$ is the gradient of $H$ (ie, $[\nabla H]_k \, = \,
\partial H / \partial z_k$), and 
\beq
  J & = & \left[\begin{array}{cc}
    0_{d \times d} & I_{d \times d} \\[7pt]
    -I_{d \times d} & 0_{d \times d} 
    \end{array}\right]
  \label{e:J}
\eeq
is a $2d \times 2d$ matrix whose quadrants are defined above in terms
identity and zero matrices.

\paragraph{Potential and kinetic energy.}
For Hamiltonian Monte Carlo, we usually use Hamiltonian functions
that can be written as follows:
\beq
   H(q,p) & = & U(q)\, + \, K(p) \label{e:HUK}
\eeq
Here, $U(q)$ is called the \textit{potential energy}, and will be
defined to be minus the log probability density of the distribution
for $q$ that we wish to sample, plus any constant that is convenient.
$K(p)$ is called the \textit{kinetic energy}, and is usually defined as
\beq
   K(p) & = & p\T M^{-1} p\, /\, 2 \label{e:quadK}
\eeq
Here, $M$ is a symmetric, positive-definite ``mass matrix'', which is 
typically diagonal, and is
often a scalar multiple of the identity matrix.  This form for
$K(p)$ corresponds to minus the log probability density (plus a constant) 
of the zero-mean Gaussian distribution with covariance matrix $M$.

With these forms for $H$ and $K$, Hamilton's equations, \eqref{e:ham-eq-a1}
and \eqref{e:ham-eq-a2}, can be written as follows, 
for $i=1,\ldots,d$:\vspace*{-6pt}
\beq
 {dq_i \over dt} & = & [M^{-1} p]_i 
   \label{e:ham-eq-b1}\\[4pt]
 {dp_i \over dt} & = & - {\partial U \over \partial q_i} 
   \label{e:ham-eq-b2}
\eeq

\vspace{-10pt}

\paragraph{A one-dimensional example.}
Consider a simple example in one dimension (for which $q$ and $p$ are
scalars and will be written without subscripts), in which the
Hamiltonian is defined as follows:\vspace*{2pt}
\beq
   H(q,p) \,=\, U(q)+K(p),\ \ \ U(q)=q^2/2,\ \ \ K(p)=p^2/2 
   \label{e:ex1}
   \\[-11pt] \nonumber
\eeq
As we'll see later in Section~\ref{ss:prob}, 
this corresponds to a Gaussian distribution for 
$q$ with mean zero and variance one.
The dynamics resulting from this Hamiltonian (following
equations~\eqref{e:ham-eq-b1} and \eqref{e:ham-eq-b2}) is\vspace*{-6pt}
\beq
  {dq \over dt}\,=\, p,\ \ \ {dp \over dt} \,=\, -q,
   \\[-10pt] \nonumber
\eeq
Solutions have the following form, for some constants $r$ and $a$:\vspace*{3pt}
\beq
  q(t)\,=\,r\cos(a+t),\ \ \ p(t) = -r\sin(a+t)
\label{e:ex1-sol}
   \\[-10pt] \nonumber
\eeq
Hence the mapping $T_s$ is a rotation by
$s$ radians clockwise around the origin in the $(q,p)$ plane.  In higher
dimensions, Hamiltonian dynamics generally does not have such a simple
periodic form, but this example does illustrate some important
properties that we will look at next.

\subsection{Properties of Hamiltonian dynamics}\label{ss:prop}

Several properties of Hamiltonian dynamics are crucial to its use in
constructing Markov chain Monte Carlo updates.  

\paragraph{Reversibility.}  First, Hamiltonian dynamics is 
\textit{reversible} --- the mapping
$T_s$ from the state at time $t$, $(q(t),p(t))$, to the state at time
$t+s$, $(q(t+s),p(t+s))$, is one-to-one, and hence has an inverse,
$T_{-s}$.  This inverse mapping is obtained by simply negating the
time derivatives in equations \eqref{e:ham-eq-a1} and
\eqref{e:ham-eq-a2}.  When the Hamiltonian has the form in equation
\eqref{e:HUK}, and $K(p)=K(-p)$, as in the quadratic form for the
kinetic energy of equation \eqref{e:quadK}, the inverse mapping can
also be obtained by negating $p$, applying $T_s$, and then negating
$p$ again.

In the simple 1D example
of equation~\eqref{e:ex1}, $T_{-s}$ is just a
counter-clockwise rotation by $s$ radians, undoing the clockwise
rotation of $T_s$.

The reversibility of Hamiltonian dynamics is important for showing
that MCMC updates that use the dynamics leave the desired distribution
invariant, since this is most easily proved by showing reversibility
of the Markov chain transitions, which requires reversibility of the
dynamics used to propose a state.

\paragraph{Conservation of the Hamiltonian.}
A second property of the dynamics is that it \textit{keeps the
Hamiltonian invariant} (ie, conserved).  This is easily seen from equations
\eqref{e:ham-eq-a1} and \eqref{e:ham-eq-a2} as follows:
\beq
  {dH \over dt} 
  & = & \sum_{i=1}^d \left[ {dq_i \over dt}{\partial H \over \partial q_i}
                      + {dp_i \over dt}{\partial H \over \partial p_i} 
        \right]
  \ \ =\ \ \sum_{i=1}^d \left[
    {\partial H \over \partial p_i}{\partial H \over \partial q_i}
            - {\partial H \over \partial q_i}{\partial H \over \partial p_i}
    \right]
  \ \ =\ \ 0
\eeq

\vspace{-10pt}

With the Hamiltonian of equation~\eqref{e:ex1}, the value of the
Hamiltonian is half the squared distance from the origin, and the
solutions (equation~\eqref{e:ex1-sol}) stay at a constant distance from 
the origin, keeping $H$ constant.

For Metropolis updates using a proposal found by Hamiltonian dynamics,
which form part of the HMC method, the acceptance probability is one
if $H$ is kept invariant.  We will see later, however, that in
practice we can only make $H$ approximately invariant, and hence we
will not quite be able to achieve this.

\paragraph{Volume preservation.}
A third fundamental property of Hamiltonian dynamics is that it
\textit{preserves volume} in $(q,p)$ space (a result known as
Liouville's Theorem).  If we apply the mapping $T_s$ to the
points in some region $R$ of $(q,p)$ space, with volume $V$, the image
of $R$ under $T_s$ will also have volume $V$.  

With the Hamiltonian of equation~\eqref{e:ex1}, the solutions
(equation~\eqref{e:ex1-sol}) are rotations, which obviously do not
change the volume.  Such rotations also do not change the shape of
a region, but this is not so in general --- Hamiltonian dynamics 
might stretch a region in one direction, as long as the region is 
squashed in some other direction so as to preserve volume.

The significance of volume preservation for MCMC is that we needn't
account for any change in volume in the acceptance probability for
Metropolis updates.  If we proposed new states using some arbitrary,
non-Hamiltonian, dynamics, we would need to compute the determinant of
the Jacobian matrix for the mapping the dynamics defines, which might
well be infeasible.

The preservation of volume by Hamiltonian dynamics can be proved in
several ways.  One is to note that the divergence of the
vector field defined by equations \eqref{e:ham-eq-a1} and
\eqref{e:ham-eq-a2} is zero, which can be seen as follows:
\beq
  \sum_{i=1}^d \left[ {\partial \over \partial q_i} {dq_i \over dt} 
                    + {\partial \over \partial p_i} {dp_i \over dt} \right]
  \,=\,
  \sum_{i=1}^d \left[ {\partial \over \partial q_i}
                      {\partial H \over \partial p_i} 
                    - {\partial \over \partial p_i} 
                      {\partial H \over \partial q_i}
  \right]  
  \,=\,
  \sum_{i=1}^d \left[   {\partial^2 H \over \partial q_i\partial p_i}
                      - {\partial^2 H \over \partial p_i\partial q_i} 
                      \right]
  \, =\, 0\ \ \ \
\eeq
A vector field with zero divergence can be shown to preserve volume
\citep{arnold:1989}.

Here, I will show informally that Hamiltonian dynamics preserves
volume more directly, without presuming this property of the
divergence.  I will, however, take as given that volume preservation
is equivalent to the determinant of the Jacobian matrix of $T_s$
having absolute value one, which is related to the well-known role of
this determinant in regard to the effect of transformations on
definite integrals and on probability density functions.

The $2d \times 2d$ Jacobian matrix of $T_s$, seen as a
mapping of $z = (q,p)$, will be written as $B_s$.  In general, $B_s$
will depend on the values of $q$ and $p$ before the mapping.  When
$B_s$ is diagonal, it is easy to see that the absolute values of its
diagonal elements are the factors by which $T_s$ stretches or
compresses a region in each dimension, so that the product of these
factors, which is equal to the absolute value of $\det(B_s)$, is the
factor by which the volume of the region changes.  I will not prove
the general result here, but note that if we were to (say) rotate the
coordinate system used, $B_s$ would no longer be diagonal, but the
determinant of $B_s$ is invariant to such transformations, and so
would still give the factor by which the volume changes.

Let's first consider volume preservation for Hamiltonian dynamics in
one dimension (ie, with $d=1$), for which we can drop the subscripts
on $p$ and $q$.  We can approximate $T_{\delta}$ for $\delta$ near
zero as follows: 
\beq
   T_{\delta}(q,p) & = & 
     \left[\begin{array}{c} q \\[6pt] p\end{array}\right] \ +\ 
     \delta \left[\begin{array}{c} dq/dt \\[6pt] dp/dt\end{array}\right] 
     \ +\ \mbox{terms of order $\delta^2$ or higher}
\eeq
Taking the time derivatives from
equations~\eqref{e:ham-eq-a1} and~\eqref{e:ham-eq-a2}, the Jacobian matrix 
can be written as\vspace{-2pt}
\beq
   B_{\delta} & = & \left[ \begin{array}{ccc} 
    \displaystyle 1 \,+\, \delta {\partial^2 H \over \partial q \partial p}
  &&\displaystyle \delta {\partial^2 H \over \partial p^2} \\[18pt]
    \displaystyle -\delta {\partial^2 H \over \partial q^2}
  &&\displaystyle  1 \,-\, \delta {\partial^2 H \over \partial p \partial q}
   \end{array}\right] 
  \ +\ \mbox{terms of order $\delta^2$ or higher}
\label{e:jacobian-1D}
\eeq
We can then write the determinant of this matrix as\vspace{2pt}
\beq
  \det(B_{\delta}) & = & 
    1 \ +\ \delta {\partial^2 H \over \partial q \partial p}
      \ -\ \delta {\partial^2 H \over \partial p \partial q}
      \ +\ \mbox{terms of order $\delta^2$ or higher} \\[6pt]
    & = &
    1 \ +\ \mbox{terms of order $\delta^2$ or higher}
\eeq
Since $\log(1+x)\approx x$ for $x$ near zero, $\log\det(B_{\delta})$
is zero except perhaps for terms of order $\delta^2$ or higher (though we will
see later that it is exactly zero).
Now consider $\log\det(B_s)$ for some time interval $s$ that is not
close to zero.  Setting $\delta=s/n$, for some integer $n$, we can
write $T_s$ as the composition of $T_{\delta}$ applied $n$ times (from
$n$ points along the trajectory), so $\det(B_s)$ is the $n$-fold product of 
$\det(B_{\delta})$ evaluated at these points.  We then find that
\beq
  \log\det(B_s) & = & \sum_{i=1}^n \log\det(B_{\delta}) \label{e:detsum} \\[6pt]
   & = & \sum_{i=1}^n \Big\{\mbox{terms of order $1/n^2$ or smaller}\Big\}
   \\[6pt]
   & = & \mbox{terms of order $1/n$ or smaller}
\eeq
Note that the value of $B_{\delta}$ in the sum in \eqref{e:detsum}
might perhaps vary with $i$, since
the values of $q$ and $p$ vary along the trajectory that produces $T_s$.
However, assuming that trajectories are not singular, the variation in 
$B_{\delta}$ must be bounded along any particular trajectory.
Taking the limit as $n\rightarrow\infty$, we conclude that $\log\det(B_s)=0$,
so $\det(B_s)=1$, and hence $T_s$ preserves volume.  

When $d>1$, the same argument applies.  The Jacobian matrix will now have
the following form (compare equation~\eqref{e:jacobian-1D}), where each
entry shown below is a $d \times d$ sub-matrix, with rows indexed by $i$
and columns by $j$:
\beq
   B_{\delta}\! & = & \!\left[ \begin{array}{ccc} 
    \displaystyle I \,+\, 
      \delta \left[{\partial^2 H \over \partial q_j \partial p_i}\right]
  &&\displaystyle 
      \delta \left[{\partial^2 H \over \partial p_j \partial p_i}\right]
  \\[24pt]
    \displaystyle 
      -\delta \left[{\partial^2 H \over \partial q_j \partial q_i}\right]
  &&\displaystyle  I \,-\, 
       \delta \left[{\partial^2 H \over \partial p_j \partial q_i}\right]
   \end{array}\right]\!
  \ +\ \mbox{terms of order $\delta^2$ or higher}\ \ \ \ \ \
\label{e:jacobian}
\eeq
As for $d=1$, the determinant of this matrix will be one plus terms of
order $\delta^2$ or higher, since all the terms of order $\delta$ cancel.
The remainder of the argument above then applies without change.

\paragraph{Symplecticness.}
Volume preservation is also a consequence of Hamiltonian dynamics
being \textit{symplectic}.  Letting $z=(q,p)$, and defining $J$
as in equation\eqref{e:J}, the symplecticness condition is that 
the Jacobian matrix, $B_s$, of the mapping $T_s$ satisfies
\beq
   B_s\T\, J^{-1}\, B_s & = & J^{-1} \label{e:symplectic}
\eeq
This implies volume conservation, since
$\det(B_s\T)\det(J^{-1})\det(B_s)\,=\, \det(J^{-1})$ implies that
$\det(B_s)^2$ is one.
When $d>1$, the symplecticness condition is stronger than volume
preservation.  Hamiltonian dynamics and the symplecticness condition can be
generalized to where $J$ is any matrix for which $J\T = -J$ and
$\det(J)\ne0$.

Crucially, reversibility, preservation of volume, and
symplecticness can be maintained exactly even when, as is necessary in
practice, Hamiltonian dynamics is approximated, as we will see next.

\subsection{Discretizing Hamilton's equations --- the leapfrog method}
\label{ss:disc}

For computer implementation, Hamilton's equations must be approximated
by discretizing time, using some small stepsize, $\varepsilon$.  Starting
with the state at time zero, we iteratively compute (approximately)
the state at times $\varepsilon$, $2\varepsilon$, $3\varepsilon$, etc.  

In discussing how to do this, I will assume that the Hamiltonian has
the form $H(q,p) =U(q)+K(p)$, as in equation~\eqref{e:HUK}.  Although
the methods below can be applied with any form for the kinetic
energy, I for simplicity assume that $K(p)=p\T M^{-1} p$, as in
equation~\eqref{e:quadK}, and furthermore that $M$ is diagonal, with
diagonal elements $m_1,\ldots,m_d$, so that
\beq
  K(p) & = & \sum_{i=1}^d {p_i^2 \over 2m_i} \label{e:indK}
\eeq

\vspace{-10pt}

\paragraph{Euler's method.}
Perhaps the best-known way to approximate the solution to a system of
differential equations is Euler's method.  For Hamilton's equations, 
this method performs the following steps, for each component
of position and momentum, indexed by $i=1,\ldots,d$:
\beq
  p_i(t+\varepsilon) & = & p_i(t) \ +\ \varepsilon\, {dp_i \over dt} (t)
    \ \ = \ \ p_i(t) \ -\ \varepsilon\, {\partial U \over \partial q_i} (q(t)) 
\label{e:euler-p}
\\[6pt]
  q_i(t+\varepsilon) & = & q_i(t) \ +\ \varepsilon\, {dq_i \over dt} (t)
                \ \ = \ \ q_i(t) \ +\ \varepsilon\, {p_i(t) \over m_i}
\label{e:euler-q}
\eeq
The time derivatives above are from the form of Hamilton's equations
given by \eqref{e:ham-eq-b1} and~\eqref{e:ham-eq-b2}.  If we start at
$t=0$ with given values for $q_i(0)$ and $p_i(0)$, we can iterate the
steps above to get a trajectory of position and momentum values at
times $\varepsilon, 2\varepsilon, 3\varepsilon, \ldots$, and hence
find (approximate) values for $q(\tau)$ and $p(\tau)$ after
$\tau/\varepsilon$ steps (assuming $\tau/\varepsilon$ is an integer).

\begin{figure}
\begin{center}
  \vspace*{-12pt}
  \includegraphics[width=6.5in]{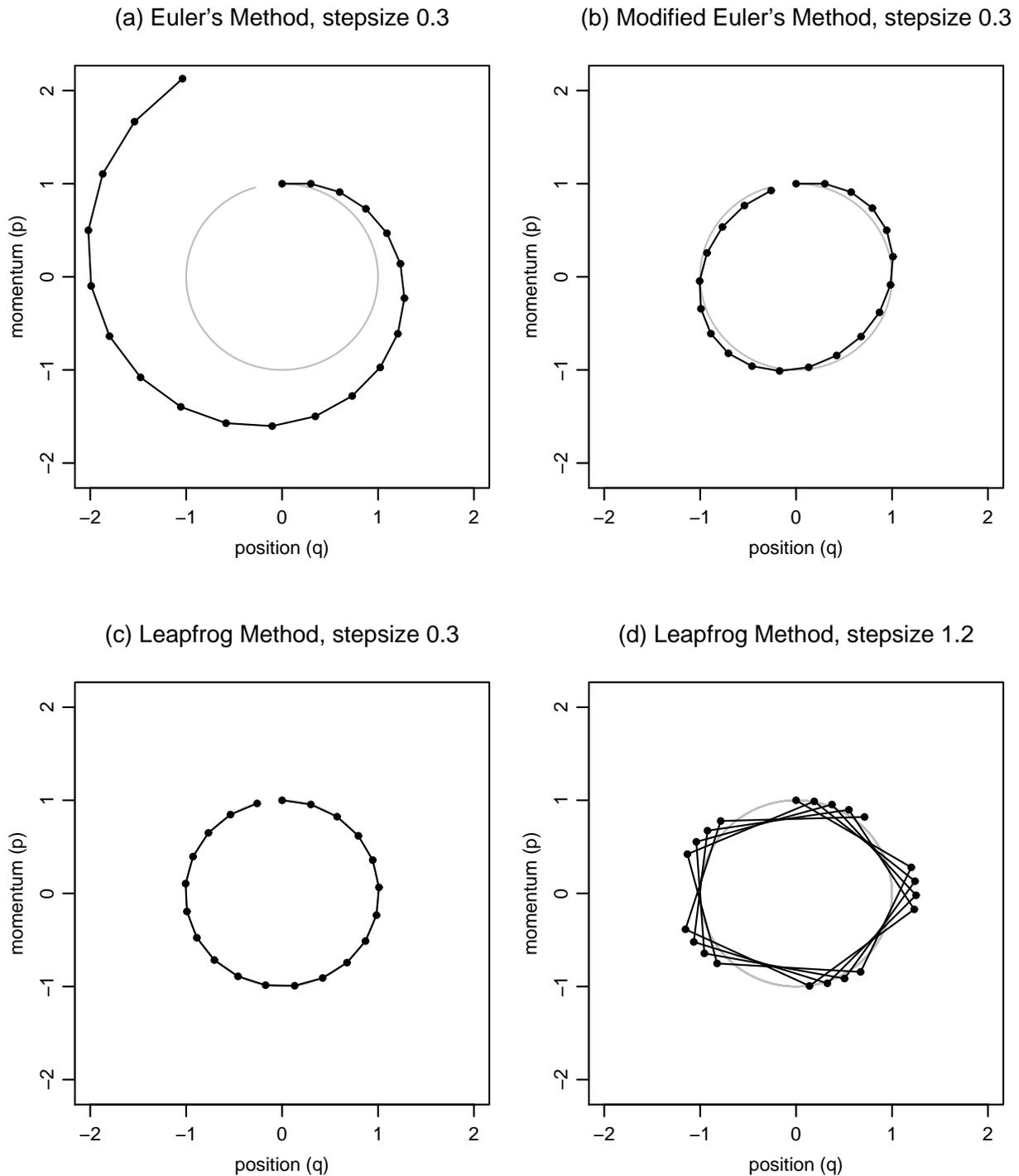}
  \vspace*{-20pt}
\end{center}
\caption{Results using three methods for approximating Hamiltonian
dynamics, when $H(q,p)=q^2/2+p^2/2$.  The initial state was $q=0$, $p=1$.
The stepsize was $\varepsilon=0.3$ for (a), (b), and (c), and $\varepsilon=1.2$
for (d).  Twenty steps of the simulated
trajectory are shown for each method, along with the true trajectory (in
gray).}\label{fig:demo}
\end{figure}

Figure~\ref{fig:demo}(a) shows the result of using Euler's method to
approximate the dynamics defined by the Hamiltonian of~\eqref{e:ex1},
starting from $q(0)=0$ and $p(0)=1$, and using a stepsize of
$\varepsilon=0.3$ for 20 steps (ie, to $\tau\,=\,0.3\times20\,=\,6$).
The results aren't good --- Euler's method produces a trajectory that
diverges to infinity, but the true trajectory is a circle.  Using a
smaller value of $\varepsilon$, and correspondingly more steps,
produces a more accurate result at $\tau=6$, but although the
divergence to infinity is slower, it is not eliminated.

\paragraph{A modification of Euler's method.}
Much better results can be obtained by slightly modifying Euler's method,
as follows:\vspace{-2pt}
\beq
  p_i(t+\varepsilon) & = & 
    p_i(t) \ -\ \varepsilon\, {\partial U \over \partial q_i} (q(t)) 
\label{e:euler2-p}
\\[6pt]
  q_i(t+\varepsilon) & = & 
    q_i(t) \ +\ \varepsilon\, {p_i(t+\varepsilon) \over m_i}
\label{e:euler2-q}
\eeq
We simply use the \textit{new} value for the momentum variables, $p_i$, when
computing the new value for the position variables, $q_i$.  A method with
similar performance can be obtained by instead updating the $q_i$ first and 
using their new values to update the $p_i$.

Figure~\ref{fig:demo}(b) shows the results using this modification of
Euler's method with $\varepsilon=0.3$.  Though not perfect, the
trajectory it produces is much closer to the true trajectory than that
obtained using Euler's method, with no tendency to diverge to
infinity.  This better performance is related to the modified method's
exact preservation of volume, which helps avoid divergence to infinity
or spiraling into the origin, since these would typically involve the
volume expanding to infinity or contracting to zero.

To see that this modification of Euler's method preserves volume
exactly despite the finite discretization of time, note that both the
transformation from $(q(t),p(t))$ to $(q(t),p(t+\varepsilon))$ via
equation~\eqref{e:euler2-p} and the transformation from
$(q(t),p(t+\varepsilon))$ to $(q(t+\varepsilon),p(t+\varepsilon))$ via
equation~\eqref{e:euler2-q} are ``shear'' transformations, in which
only some of the variables change (either the $p_i$ or the $q_i$), by
amounts that depend only on the variables that do not change.  Any
shear transformation will preserve volume, since its Jacobian matrix
will have determinant one (as the only non-zero term in the
determinant will be the product of diagonal elements, which will all
be one).

\paragraph{The leapfrog method.}
Even better results can be obtained with the \textit{leapfrog} method,
which works as follows:
\beq
  p_i(t+\varepsilon/2) & = &
    p_i(t) \ -\ (\varepsilon/2)\, {\partial U \over \partial q_i} (q(t)) 
\label{e:leap1}\\[6pt]
  q_i(t+\varepsilon) & = & 
    q_i(t) \ +\ \varepsilon\, {p_i (t+\varepsilon/2) \over m_i} 
\label{e:leap2}\\[6pt]
  p_i(t+\varepsilon) & = &
    p_i(t+\varepsilon/2) \ -\ (\varepsilon/2)\, {\partial U \over \partial q_i} 
    (q(t+\varepsilon))
\label{e:leap3}
\eeq
We start with a half step for the momentum variables, then do a
full step for the position variables, using the new values of the
momentum variables, and finally do another half step for the momentum
variables, using the new values for the position variables.  An
analogous scheme can be used with any kinetic energy function, with
$\partial K/\partial p_i$ replacing $p_i/m_i$ above.


When we apply equations~\eqref{e:leap1} to~\eqref{e:leap3} a second
time to go from time $t+\varepsilon$ to $t+2\varepsilon$, we can
combine the last half step of the first update, from
$p_i(t+\varepsilon/2)$ to $p_i(t+\varepsilon)$, with the first half
step of the second update, from $p_i(t+\varepsilon)$ to
$p_i(t+\varepsilon+\varepsilon/2)$.  The leapfrog
method then looks very similar to the modification of Euler's method in
equations~\eqref{e:euler2-q} and~\eqref{e:euler2-p}, 
except that leapfrog performs half steps for momentum at the very
beginning and very end of the trajectory, and the time labels of the
momentum values computed are shifted by $\varepsilon/2$.

The leapfrog method preserves volume exactly, since each of
\eqref{e:leap1} to \eqref{e:leap3} are shear transformations.  Due to
its symmetry, it is also reversible by simply negating $p$, applying
the same number of steps again, and then negating $p$ again.

Figure~\ref{fig:demo}(c) shows the results using the leapfrog method
with a stepsize of $\varepsilon=0.3$, which are indistinguishable from
the true trajectory, at the scale of this plot.  In
Figure~\ref{fig:demo}(d), the results of using the leapfrog method
with $\varepsilon=1.2$ are shown (still with 20 steps, so almost four
cycles are seen, rather than almost one).  With this larger stepsize,
the approximation error is clearly visible, but the trajectory still
remains stable (and will stay stable indefinitely).  Only when the
stepsize approaches $\varepsilon=2$ do the trajectories become
unstable. 

\paragraph{Local and global error of discretization methods.}

\mbox{\hspace{-6pt}I} will briefly discuss how the error\linebreak
from discretizing the dynamics behaves in the limit as the stepsize,
$\varepsilon$, goes to zero; Leimkuhler and Reich (2004) provide a much
more detailed discussion.  For useful methods, the error goes to zero
as $\varepsilon$ goes to zero, so that any upper limit on the error
will apply (apart from a usually unknown constant factor) to any
differentiable function of state --- eg, if the error for $(q,p)$ is
no more than order $\varepsilon^2$, the error for $H(q,p)$ will also
be no more than \mbox{order~$\varepsilon^2$.}

The \textit{local error} is the error after one step, that moves from
time $t$ to time $t+\varepsilon$.  The \textit{global error} is the
error after simulating for some fixed time interval, $s$, which will
require $s/\varepsilon$ steps.  If the local error is order
$\varepsilon^p$, the global error will be order $\varepsilon^{p-1}$
--- the local errors of order $\varepsilon^p$ accumulate over the
$s/\varepsilon$ steps to give an error of order $\varepsilon^{p-1}$.
If we instead fix $\varepsilon$ and consider increasing the time, $s$,
for which the trajectory is simulated, the error can in general
increase exponentially with $s$.  Interestingly, however, this is
often not what happens when simulating Hamiltonian dynamics with a
symplectic method, as can be seen in Figure~\ref{fig:demo}.

The Euler method and its modification above have order $\varepsilon^2$
local error and order $\varepsilon$ global error.  The leapfrog method
has order $\varepsilon^3$ local error and order $\varepsilon^2$ global
error.  As shown by \citet[Section 4.3.3]{leimkuhler:2004} this
difference is a consequence of leapfrog being reversible, since any
reversible method must have global error that is of even order in
$\varepsilon$.

\section{MCMC from Hamiltonian dynamics}\label{s:mcmc}

Using Hamiltonian dynamics to sample from a distribution requires
translating the density function for this distribution to a potential
energy function and introducing ``momentum'' variables to go with the
original variables of interest (now seen as ``position'' variables).
We can then simulate a Markov chain in which each iteration resamples the
momentum and then does a Metropolis update with a proposal found using
Hamiltonian dynamics.

\subsection{Probability and the Hamiltonian --- canonical distributions}
\label{ss:prob}

The distribution we wish to sample can be related to a potential
energy function via the concept of a \textit{canonical distribution}
from statistical mechanics.  Given some energy function, $E(x)$, for
the state, $x$, of some physical system, the canonical distribution
over states has probability or probability density function
\beq
   P(x) & = & {1 \over Z}\, \exp(-E(x)/T)\label{e:canT}
\eeq
Here, $T$ is the temperature of the system\footnote{Note to physicists:\ \
I assume here that temperature is measured in units that make
Boltzmann's constant be one.}, and $Z$ is the normalizing
constant needed for this function to sum or integrate to one.  Viewing this
the opposite way, if we are interested in some distribution with
density function $P(x)$, we can obtain it as a canonical distribution
with $T=1$ by setting $E(x) = -\log P(x) - \log Z$, where $Z$ is
any convenient positive constant.

The Hamiltonian is an energy function for the joint state of
``position'', $q$, and ``momentum'', $p$, and so defines a joint
distribution for them, as follows:
\beq
   P(q,p) & = & {1 \over Z}\, \exp(-H(q,p)/T)
\eeq
Note that the invariance of $H$ under Hamiltonian dynamics means that 
a Hamiltonian trajectory will (if simulated exactly) move within a 
hyper-surface of constant probability density.

If $H(q,p) = U(q) + K(p)$, the joint density is
\beq
   P(q,p) & = & {1 \over Z}\, \exp(-U(q)/T)\, \exp(-K(p)/T)
\label{e:Hcan}
\eeq
and we see that $q$ and $p$ are independent, and each have canonical
distributions, with energy functions $U(q)$ and $K(p)$.  We will use 
$q$ to represent the variables of interest, and introduce $p$ just
to allow Hamiltonian dynamics to operate.

In Bayesian statistics, the posterior distribution for the model
parameters is the usual focus of interest, and hence these parameters
will take the role of the position, $q$.  We can express the posterior
distribution as a canonical distribution (with $T=1$) using a
potential energy function defined as follows:\vspace*{-2pt}
\beq
  U(q) & = & -\log \Big[\pi(q)L(q|D)\Big]
\eeq
where $\pi(q)$ is the prior density, and $L(q|D)$ is the likelihood 
function given data $D$.

\subsection{The Hamiltonian Monte Carlo algorithm}\label{ss:HMC}

We now have the background needed to present the Hamiltonian Monte
Carlo (HMC) algorithm.  HMC can be used to sample only from continuous
distributions on $R^d$ for which the density function can be evaluated
(perhaps up to an unknown normalizing constant). For the moment, I
will also assume that the density is non-zero everywhere (but this is
relaxed in Section~\ref{ss:split}).  We must also be able to compute
the partial derivatives of the log of the density function.  These
derivatives must therefore exist, except perhaps on a set of points
with probability zero, for which some arbitrary value could be
returned.

HMC samples from the canonical distribution for $q$ and $p$ defined by
equation~\eqref{e:Hcan}, in which $q$ has the distribution of
interest, as specified using the potential energy function $U(q)$.  We
can choose the distribution of the momentum variables, $p$, which are
independent of $q$, as we wish, specifying the distribution via the
kinetic energy function, $K(p)$.  Current practice with HMC is to use
a quadratic kinetic energy, as in equation~\eqref{e:quadK}, which
leads $p$ to have a zero-mean multivariate Gaussian distribution.
Most often, the components of $p$ are specified to be independent,
with component $i$ having variance $m_i$.  The kinetic energy function
producing this distribution (setting $T=1$) is
\beq
  K(p) & = & \sum_{i=1}^d {p_i^2 \over 2m_i}
\label{e:Km}
\eeq
We will see in Section~\ref{s:pr-th} how the choice for 
the $m_i$ affects performance.

\paragraph{The two steps of the HMC algorithm.}
Each iteration of the HMC algorithm has two steps.  The first 
changes only the momentum; the second may change both position and
momentum.  Both steps leave the canonical joint distribution of
$(q,p)$ invariant, and hence their combination also leaves this
distribution invariant.

In the first step, new values for the momentum variables are randomly
drawn from their Gaussian distribution, independently of the current
values of the position variables.  For the kinetic energy of
equation~\eqref{e:Km}, the $d$ momentum variables are independent,
with $p_i$ having mean zero and variance $m_i$.  Since $q$ isn't
changed, and $p$ is drawn from it's correct conditional distribution
given $q$ (the same as its marginal distribution, due to independence),
this step obviously leaves the canonical joint distribution invariant.

In the second step, a Metropolis update is performed, using
Hamiltonian dynamics to propose a new state.  Starting with the
current state, $(q,p)$, Hamiltonian dynamics is simulated for $L$ 
steps using the Leapfrog method (or some other
reversible method that preserves volume), with a stepsize
of $\varepsilon$. Here, $L$ and $\varepsilon$ are parameters of the
algorithm, which need to be tuned to obtain good performance (as
discussed below in Section~\ref{ss:tuning}).  The momentum variables
at the end of this $L$-step trajectory are then negated, giving a
proposed state $(q^*,p^*)$.  This proposed state is accepted as the
next state of the Markov chain with probability 
\beq
     \min\Big[ 1,\, \exp(-H(q^*,p^*)+H(q,p)) \Big]
 & \!=\! & \min\Big[ 1,\, \exp(-U(q^*)+U(q)-K(p^*)+K(p)) \Big]\ \ \ \ \ \ \ \
\label{e:HMC-accept}
\eeq
If the proposed state is not accepted (ie, it is rejected), the next
state is the same as the current state (and is counted again when
estimating the expectation of some function of state by its average
over states of the Markov chain).  The negation of the momentum variables
at the end of the trajectory makes the Metropolis proposal symmetrical,
as needed for the acceptance probability above to be valid.
This negation need not be done in practice,
since $K(p)=K(-p)$, and the momentum will be replaced before it is
used again, in the first step of the next iteration.  (This assumes
that these HMC updates are the only ones performed.)

If we look at HMC as sampling from the joint distribution of $q$ and
$p$, the Metropolis step using a proposal found by Hamiltonian
dynamics leaves the probability density for $(q,p)$ unchanged or
almost unchanged.  Movement to $(q,p)$ points with a different
probability density is accomplished only by the first step in an HMC
iteration, in which $p$ is replaced by a new value.  Fortunately, this
replacement of $p$ can change the probability density for $(q,p)$ by a
large amount, so movement to points with a different probability
density is not a problem (at least not for this reason).  Looked at in
terms of $q$ only, Hamiltonian dynamics for $(q,p)$ can produce a
value for $q$ with a much different probability density (equivalently,
a much different potential energy, $U(q)$).  However, the resampling
of the momentum variables is still crucial to obtaining the proper
distribution for $q$.  Without resampling, $H(q,p) = U(q) + K(p)$ will
be (nearly) constant, and since $K(p)$ and $U(q)$ are non-negative,
$U(q)$ could never exceed the initial value of $H(q,p)$ if no
resampling for $p$ were done.

\begin{figure}

{ \begin{verbatim}
  HMC = function (U, grad_U, epsilon, L, current_q)
  {
    q = current_q
    p = rnorm(length(q),0,1)  # independent standard normal variates
    current_p = p
  
    # Make a half step for momentum at the beginning
  
    p = p - epsilon * grad_U(q) / 2  
  
    # Alternate full steps for position and momentum
  
    for (i in 1:L)
    { 
      # Make a full step for the position
  
      q = q + epsilon * p     
  
      # Make a full step for the momentum, except at end of trajectory
  
      if (i!=L) p = p - epsilon * grad_U(q)
    }
  
    # Make a half step for momentum at the end.
  
    p = p - epsilon * grad_U(q) / 2  
  
    # Negate momentum at end of trajectory to make the proposal symmetric
  
    p = -p
  
    # Evaluate potential and kinetic energies at start and end of trajectory
  
    current_U = U(current_q)
    current_K = sum(current_p^2) / 2
    proposed_U = U(q)
    proposed_K = sum(p^2) / 2
  
    # Accept or reject the state at end of trajectory, returning either
    # the position at the end of the trajectory or the initial position
  
    if (runif(1) < exp(current_U-proposed_U+current_K-proposed_K)) 
    { 
      return (q)  # accept
    }
    else
    { 
      return (current_q)  # reject
    }
  }
\end{verbatim}}\vspace*{-4pt}

\caption{The Hamiltonian Monte Carlo algorithm}\label{f:HMC}

\end{figure}

A function that implements a single iteration of the HMC algorithm,
written in the R language\footnote{R is available for free from
\texttt{r-project.org}}\!\!,\, is shown in Figure~\ref{f:HMC}.  Its
first two arguments are functions --- \texttt{U}, which returns the
potential energy given a value for $q$, and \texttt{grad\_U}, which
returns the vector of partial derivatives of $U$ given $q$.  
Other arguments are the stepsize, \texttt{epsilon}, for leapfrog
steps, the number of leapfrog steps in the trajectory, \texttt{L},
and the current position, \texttt{current\_q}, that the trajectory
starts from.  Momentum variables are sampled within this function, and
discarded at the end, with only the next position being returned.  The
kinetic energy is assumed to have the simplest form, $K(p) = \sum
p_i^2/2$ (ie, all $m_i$ are one).  In this program, all components of
$p$ and of $q$ are updated simultaneously, using vector operations.
This simple implementation of HMC is available from my web
page\footnote{At \texttt{www.cs.utoronto.ca/$\sim$radford}}\!\!,\,
along with other R programs with extra
features helpful for practical use, and that illustrate some of the
variants of HMC in Section~\ref{s:var}.

\paragraph{Proof that HMC leaves the canonical distribution invariant.}
The Metropolis update above is reversible with respect to the canonical
distribution for $q$ and $p$ (with $T=1$), a condition also known as ``detailed
balance'', and which can be phrased informally as follows.  Suppose we
partition the $(q,p)$ space into regions $A_k$, each with the same small
volume $V$.  Let the image of $A_k$ with respect to the operation of
$L$ leapfrog steps, plus a negation of the momentum, be $B_k$.  Due to
the reversibility of the leapfrog steps, the $B_k$ will also partition
the space, and since the leapfrog steps preserve volume (as does
negation), each $B_k$ will also have volume $V$.  Detailed balance
holds if, for all $i$ and $j$,
\beq
   P(A_i)T(B_j|A_i) & = & P(B_j)T(A_i|B_j)
\label{e:balance1}
\eeq
where $P$ is probability under the canonical distribution, and
$T(X|Y)$ is the conditional probability of proposing and then
accepting a move to region $X$ 
if the current state is in region $Y$.  Clearly, when $i \ne j$,
$T(A_i|B_j)=T(B_j|A_i)=0$ and so equation~\eqref{e:balance1} will be
satisfied.  Since the Hamiltonian is continuous almost everywhere, in
the limit as the regions $A_k$ and $B_k$ become smaller, the 
Hamiltonian becomes effectively constant within each region, with
value $H_X$ in region $X$, and hence the canonical probability 
density and the transition probabilities become effectively constant 
within each region as well.  We can now rewrite 
equation~\eqref{e:balance1} for $i=j$ (say both equal to $k$) as\vspace{2pt}
\beq
   {V \over Z}\exp(-H_{A_k}) \min\!\Big[1,\,\exp(-H_{B_k}\!\!+\!H_{A_k})\Big] 
   & \!=\! & 
   {V \over Z}\exp(-H_{B_k}) \min\!\Big[1,\,\exp(-H_{A_k}\!\!+\!H_{B_k})\Big]
   \ \ \ \ \ \ \ \
\label{e:balancew}
\eeq\vspace{2pt}%
which is easily seen to be true.

Detailed balance implies that this Metropolis update leaves the canonical
distribution for $q$ and $p$ invariant.  This can be seen as follows.
Let $R(X)$ be the probability that the Metropolis update for a state
in the small region $X$ leads to rejection of the proposed state.  
Suppose that the current
state is distributed according to the canonical distribution.  
The probability that the next state is in a small region $B_k$ is the sum of
the probability that the current state is in $B_k$ and the update leads
to rejection, and the probability that the current
state is in some region from which a move to $B_k$ is proposed and accepted.
The probability of the next state being in $B_k$ can therefore be written as
\beq
    P(B_k)R(B_k) \ +\ \sum_i P(A_i)T(B_k|A_i) 
    & = & P(B_k)R(B_k) \ +\ \sum_i P(B_k)T(A_i|B_k) \\[3pt]
    & = & P(B_k)R(B_k) \ +\ P(B_k) \sum_i T(A_i|B_k)\ \ \ \ \\[3pt]
    & = & P(B_k)R(B_k) \ +\ P(B_k) (1-R(B_k)) \\[8pt]
    & = & P(B_k)
\eeq
The Metropolis update within HMC therefore leaves the canonical 
distribution invariant.  

Since both the sampling of momentum variables and the Metropolis
update with a proposal found by Hamiltonian dynamics leave the
canonical distribution invariant, the HMC algorithm as a whole does as
well.  

\paragraph{Ergodicity of HMC.}
Typically, the HMC algorithm will also be ``ergodic'' --- it will not
be trapped in some subset of the state space, and hence will asymptotically
converge to its (unique) invariant distribution.  In an HMC iteration,
any value can be sampled for the momentum variables, which can
typically then affect the position variables in arbitrary ways.
However, ergodicity can fail if the $L$ leapfrog steps in a trajectory
produce an exact periodicity for some function of state.  For example,
with the simple Hamiltonian of equation~\eqref{e:ex1}, the exact
solutions (given by equation~\eqref{e:ex1-sol}) are periodic with
period $2\pi$.  Approximate trajectories found with $L$ leapfrog steps
with stepsize $\varepsilon$ may return to the same position coordinate
when $L\varepsilon$ is approximately $2\pi$.  HMC with such values for
$L$ and $\varepsilon$ will not be ergodic.  For nearby values of $L$
and $\varepsilon$, HMC may be theoretically ergodic, but take a very
long time to move about the full state space.  


This potential problem of non-ergodicity can be solved by randomly
choosing $\varepsilon$ or $L$ (or both) from some fairly small
interval \citep{mackenzie:1989}.  Doing this routinely may be
advisable. Although in real problems interactions between variables
typically prevent any exact periodicities from occurring, near
periodicities might still slow HMC considerably.

\subsection{Illustrations of HMC and its benefits}\label{ss:HMC-ill}

Here, I will illustrate some practical issues with HMC, and
demonstrate its potential to sample much more efficiently than simple
methods such as random-walk Metropolis.  I use simple Gaussian
distributions for these demonstrations, so that the results can be
compared with known values, but of course HMC is typically used for
more complex distributions.

\paragraph{Trajectories for a two-dimensional problem.}
Consider sampling from a distribution for two variables
that is bivariate Gaussian, with means of zero, standard deviations 
of one, and correlation 0.95.  We regard these as ``position'' variables,
and introduce two corresponding ``momentum'' variables, defined to have a
Gaussian distribution
with means of zero, standard deviations of one, and zero correlation.  
We then define the Hamiltonian as\vspace*{-4pt}
\beq
  H(q,p) & = & q\T \Sigma^{-1} q \,/\, 2 \ + \ p\T p\, /\, 2, \ \ \ \ \
  \mbox{with}\ \ 
  \Sigma=\left[\begin{array}{cc} 1 & 0.95 \\[4pt] 0.95 & 1 \end{array}\right]
\eeq

\begin{figure}[t]
\begin{center}
  \vspace*{-12pt}
  \includegraphics[height=2.3in,width=6.5in]{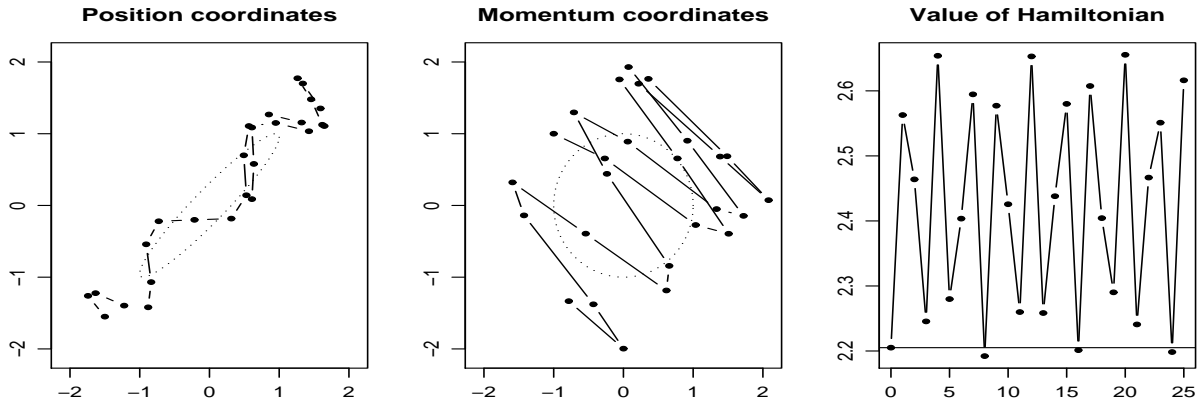}
  \vspace*{-35pt}
\end{center}\caption{A trajectory for a 2D Gaussian distribution,
simulated using 25 leapfrog steps with a stepsize of 0.25.  The
ellipses plotted are one standard deviation from the means.  The
initial state had $q = [-1.50, -1.55]\T$ and $p=[-1,1]\T$.
}\label{fig:demo2d-a}
\end{figure}

Figure~\ref{fig:demo2d-a} shows a trajectory based on this
Hamiltonian, such as might be used to propose a new state in the
Hamiltonian Monte Carlo method, computed using $L=25$
leapfrog steps, with a stepsize of $\varepsilon=0.25$.  Since the full
state space is four-dimensional, the Figure shows the two position
coordinates and the two momentum coordinates in separate plots, while
the third plot shows the value of the Hamiltonian after each leapfrog
step.

Notice that this trajectory does not resemble a random-walk.  Instead,
starting from the lower-left corner, the position variables
systematically move upwards and to the right, until they reach the
upper-right corner, at which point the direction of motion is
reversed.  The consistency of this motion results from the role of the
momentum variables.  The projection of $p$ in the diagonal direction
will change only slowly, since the gradient in that direction is
small, and hence the direction of diagonal motion stays the same for
many leapfrog steps.  While this large-scale diagonal motion is
happening, smaller-scale oscillations occur, moving back and forth
across the ``valley'' created by the high correlation between the
variables.

The need to keep these smaller oscillations under control limits the
stepsize that can be used.  As can be seen in the rightmost plot in
Figure~\ref{fig:demo2d-a}, there are also oscillations in the value of
the Hamiltonian (which would be constant if the trajectory were
simulated exactly).  If a larger stepsize were used, these
oscillations would be larger.  At a critical stepsize
($\varepsilon=0.45$ in this example), the trajectory becomes unstable,
and the value of the Hamiltonian grows without bound.  As long as the
stepsize is less than this, however, the error in the Hamiltonian
stays bounded regardless of the number of leapfrog steps done.  This
lack of growth in the error is not guaranteed for all Hamiltonians,
but it does hold for many distributions more complex than Gaussians.
As can be seen, however, the error in the Hamiltonian along the
trajectory does tend to be positive more often than negative.  In this
example, the error is $+0.41$ at the end of the trajectory, so if this
trajectory were used for an HMC proposal, the probability of accepting
the end-point as the next state would be $\exp(-0.41) = 0.66$.

\begin{figure}[p]
\begin{center}
  \vspace*{-12pt}
  \includegraphics[width=6.75in]{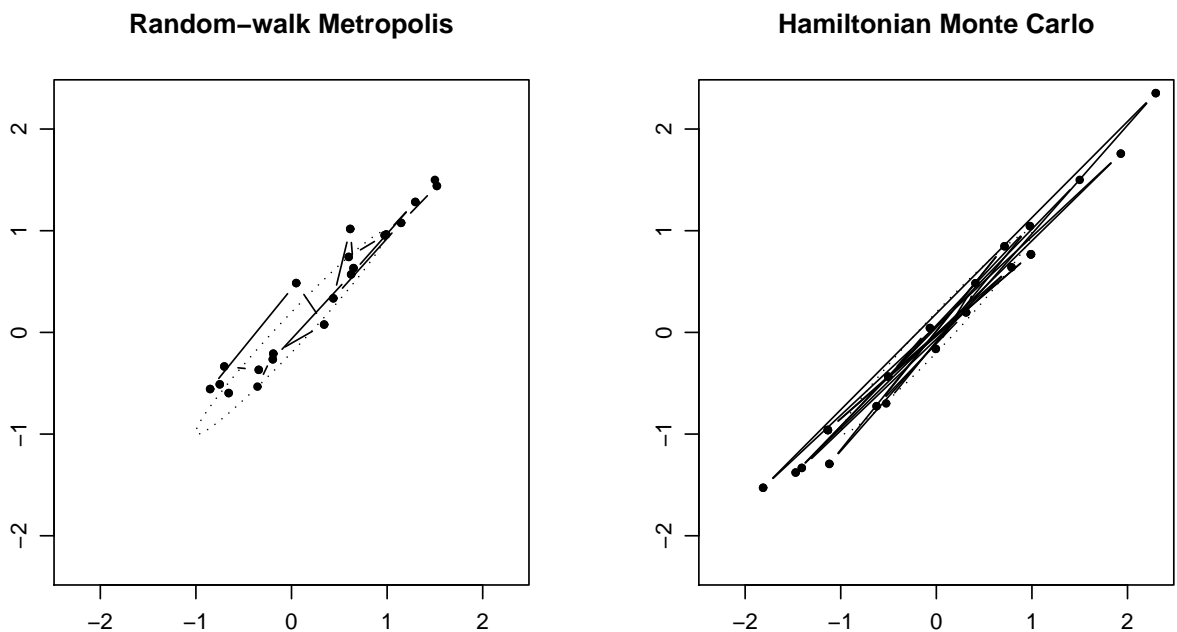}
  \vspace*{-55pt}
\end{center}
\caption{Twenty iterations of the random-walk Metropolis method (with 20 updates
per iteration) and of the Hamiltonian Monte Carlo method (with 20 leapfrog steps
per trajectory) for a 2D Gaussian distribution with marginal standard deviations
of one and correlation 0.98.  Only the two position coordinates are plotted,
with ellipses drawn one standard deviation away from the 
mean.}\label{fig:demo2d-b}
\end{figure}

\begin{figure}[p]
\begin{center}
  \vspace*{-12pt}
  \includegraphics[width=6.75in]{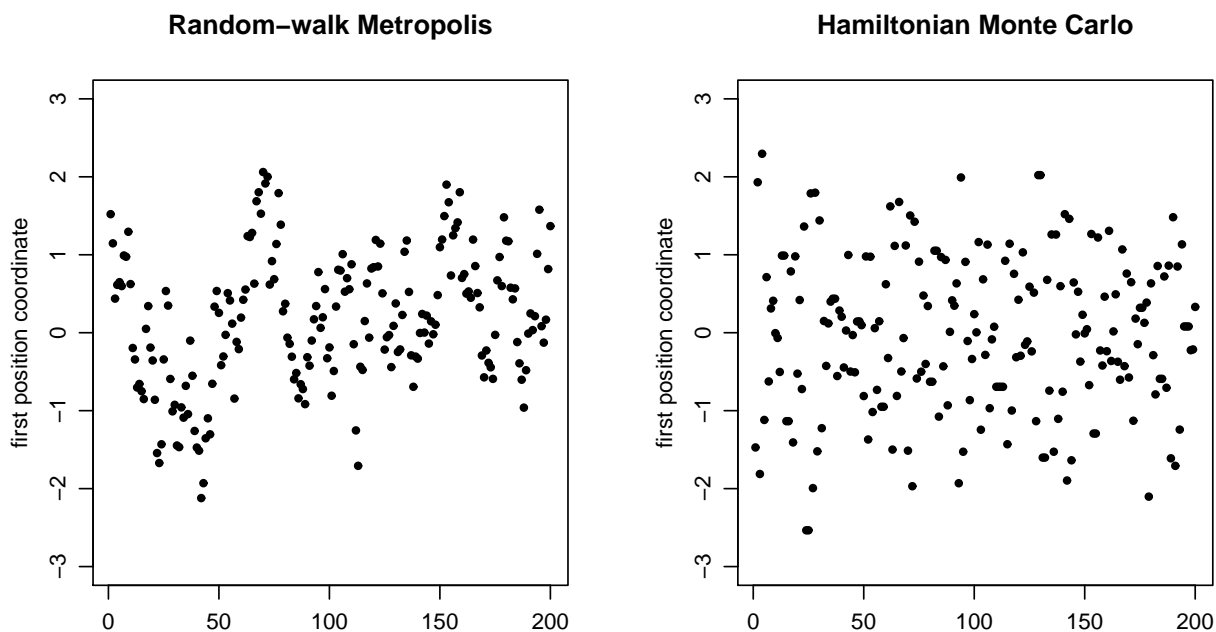}
  \vspace*{-55pt}
\end{center}
\caption{Two hundred iterations, starting with the twenty iterations shown
above, with only the first position coordinate plotted.}\label{fig:demo2d-c}
\end{figure}

\paragraph{Sampling from a two-dimensional distribution.}  Figures
\ref{fig:demo2d-b} and \ref{fig:demo2d-c} show the results of using
HMC and a simple random-walk Metropolis method to sample from a
bivariate Gaussian similar to the one just discussed, but with
stronger correlation of 0.98.

In this example, as in the previous one, HMC used a kinetic energy
(defining the momentum distribution) of $K(p) = p\T p/2$.  The results
of 20 HMC iterations, using trajectories of $L=20$ leapfrog steps with
stepsize $\varepsilon=0.18$ are shown in the right plot of
Figure~\ref{fig:demo2d-b}.  These values were chosen so that the trajectory
length, $\varepsilon L$, is sufficient to move to a distant point in the
distribution, without being so large that the trajectory will often
waste computation time by doubling back on itself.  The rejection rate
for these trajectories was 0.09.

Figure~\ref{fig:demo2d-b} also shows every 20th state from 400
iterations of random-walk Metropolis, with a bivariate Gaussian
proposal distribution with the current state as mean, zero
correlation, and the same standard deviation for the two coordinates.
The standard deviation of the proposals for this example was 0.18,
which is the same as the stepsize used for HMC proposals, so that the
change in state in these random-walk proposals was comparable to that
for a single leapfrog step for HMC.  The rejection rate for these
random-walk proposals was 0.37.

One can see in Figure~\ref{fig:demo2d-b} how the systematic motion
during an HMC trajectory (illustrated in Figure~\ref{fig:demo2d-a})
produces larger changes in state than a corresponding number of
random-walk Metropolis iterations.  Figure~\ref{fig:demo2d-c}
illustrates this difference for longer runs of $20\times200$
random-walk Metropolis iterations and of 200 HMC iterations.

\paragraph{The benefit of avoiding random walks.}  Avoidance of
random-walk behaviour, as illustrated above, is one major benefit of
Hamiltonian Monte Carlo.  In this example, because of the high
correlation between the two position variables, keeping the acceptance
probability for random-walk Metropolis reasonably high requires that
the changes proposed have a magnitude comparable to the standard
deviation in the most constrained direction (0.14 in this example, the
square root of the smallest eigenvalue of the covariance matrix).  The
changes produced using one Gibbs sampling scan would be of similar
magnitude.  The number of iterations needed to reach a state almost
independent of the current state is mostly determined by how long it
takes to explore the less constrained direction, which for this
example has standard deviation 1.41 --- about ten times greater than
the standard deviation in the most constrained direction.  We might
therefore expect that we would need around ten iterations of
random-walk Metropolis in which the proposal was accepted to move to a
nearly independent state.  But the number needed is actually roughly
the square of this --- around 100 iterations with accepted proposals
--- because the random-walk Metropolis proposals have no tendency to
move consistently in the same direction.

To see this, note that the variance of the position after $n$ 
iterations of random walk Metropolis from some start state will
grow in proportion to $n$ (until this variance becomes comparable to
the overall variance of the state), since the position is the sum of
mostly independent movements for each iteration.  The \textit{standard
deviation} of the amount moved (which gives the typical amount of
movement) is therefore proportional to $\sqrt{n}$.

The stepsize used for the leapfrog steps is similarly limited by the
most constrained direction, but the movement will be in the same
direction for many steps.  The distance moved after $n$ steps will
therefore tend to be proportional to $n$, until the distance moved
becomes comparable to the overall width of the distribution.  The
advantage compared to movement by a random walk will be a factor
roughly equal to the ratio of the standard deviations in the least
confined direction and most confined direction --- about 10 here.

Because avoiding a random walk is so beneficial, the optimal standard
deviation for random-walk Metropolis proposals in this example is
actually much larger than the value of 0.18 used here.  A proposal
standard deviation of 2.0 gives a very low acceptance rate (0.06), but
this is more than compensated for by the large movement (to a nearly
independent point) on the rare occasions when a proposal is accepted,
producing a method that is about as efficient as HMC.  However, this
strategy of making large changes with a small acceptance rate works
only when, as here, the distribution is tightly constrained in only
one direction.

\paragraph{Sampling from a 100-dimensional distribution.}
More typical behaviour of HMC and
random-walk Metropolis is illustrated by a 100-dimensional
multivariate Gaussian distribution in which the variables are
independent, with means of zero, and standard deviations of $0.01,\,
0.02,\,\ldots,\,0.99,\,1.00$.  Suppose we have no
knowledge of the details of this distribution, so we will use HMC with
the same simple, rotationally symmetric kinetic energy function as
above, $K(p) = p\T p / 2$, and use random-walk Metropolis proposals in
which changes to each variable are independent, all with the same
standard deviation.  As discussed below in Section~\ref{ss:linear},
the performance of both these sampling methods is invariant to
rotation, so this example is illustrative of how they perform on any
multivariate Gaussian distribution in which the square roots of the
eigenvalues of the covariance matrix are
$0.01,\,0.02,\,\ldots,\,0.99,\,1.00$.

For this problem, the position coordinates, $q_i$, and corresponding
momentum coordinates, $p_i$, are all independent, so the
leapfrog steps used to simulate a trajectory operate independently for
each $(q_i,p_i)$ pair.  However, whether the trajectory is accepted
depends on the total error in the Hamiltonian due to the leapfrog
discretization, which is a sum of the errors due to each $(q_i,p_i)$
pair (for the terms in the Hamiltonian involving this pair).  Keeping
this error small requires limiting the leapfrog stepsize to a value
roughly equal to the smallest of the standard deviations (0.01), which
implies that many leapfrog steps will be needed to move a distance
comparable to the largest of the standard deviations (1.00).

Consistent with this, I applied HMC to this distribution using
trajectories with \mbox{$L=150$} and with $\varepsilon$ 
randomly selected for each iteration, uniformly from $(0.0104,0.0156)$, which
is $0.013 \pm 20\%$.  I used random-walk Metropolis with proposal
standard deviation drawn uniformly from $(0.0176,0.0264)$, which is
$0.022 \pm 20\%$.  These are close to optimal settings for both
methods.  The rejection rate was 0.13 for HMC and 0.75 for random-walk
Metropolis.

\begin{figure}[t]
\begin{center}
  \vspace*{-12pt}
  \includegraphics[width=5.5in]{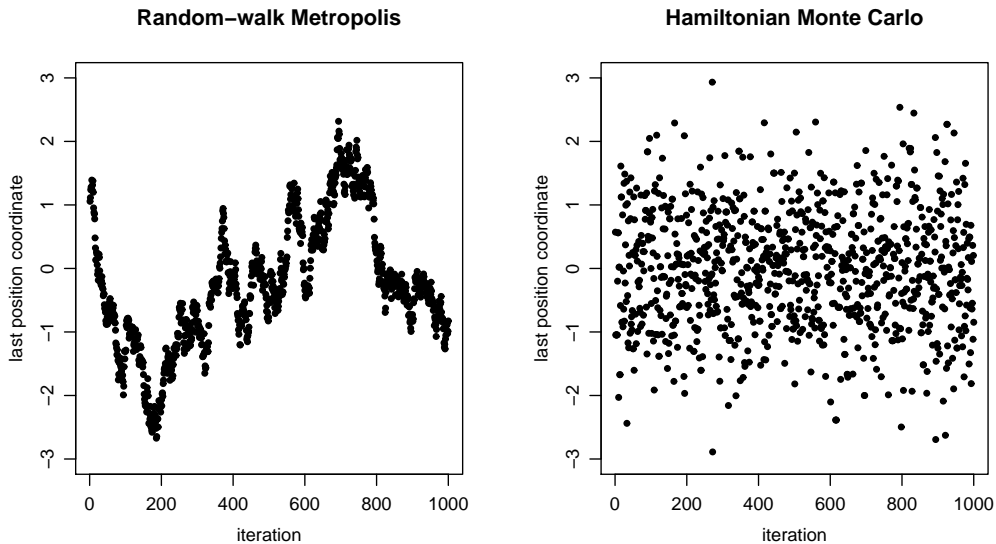}
  \vspace*{-25pt}
\end{center}
\caption{Values for the variable with largest standard deviation
for the 100-dimensional example, from a random-walk Metropolis run
and an HMC run with $L=150$.  To match computation time, 150 
updates were counted 
as one iteration for random-walk Metropolis.}\label{fig:demo100d-trace}
\end{figure}

\begin{figure}[p]
\begin{center}
  \vspace*{-12pt}
  \includegraphics[width=6.75in]{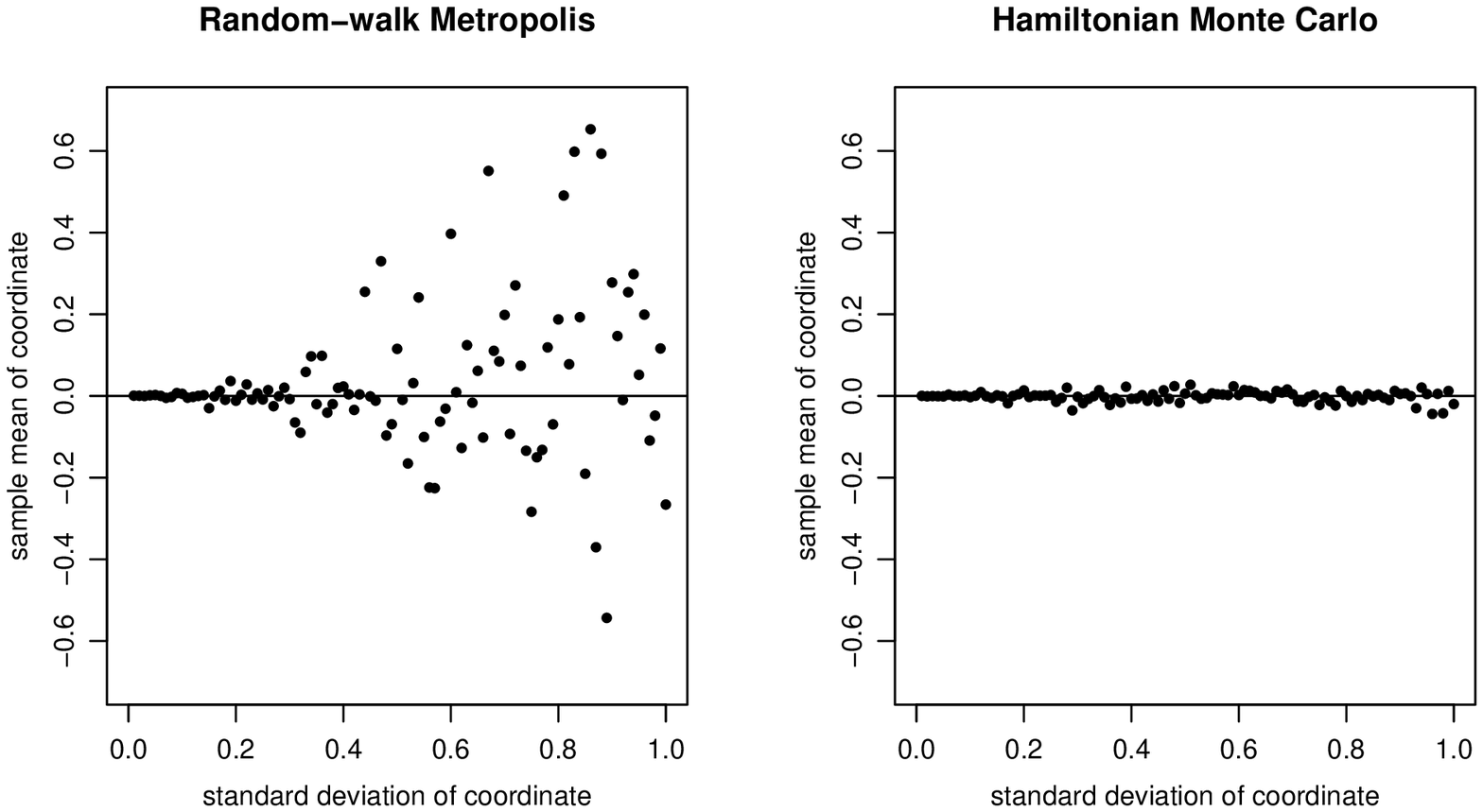}

  \vspace*{-25pt}

  \includegraphics[width=6.75in]{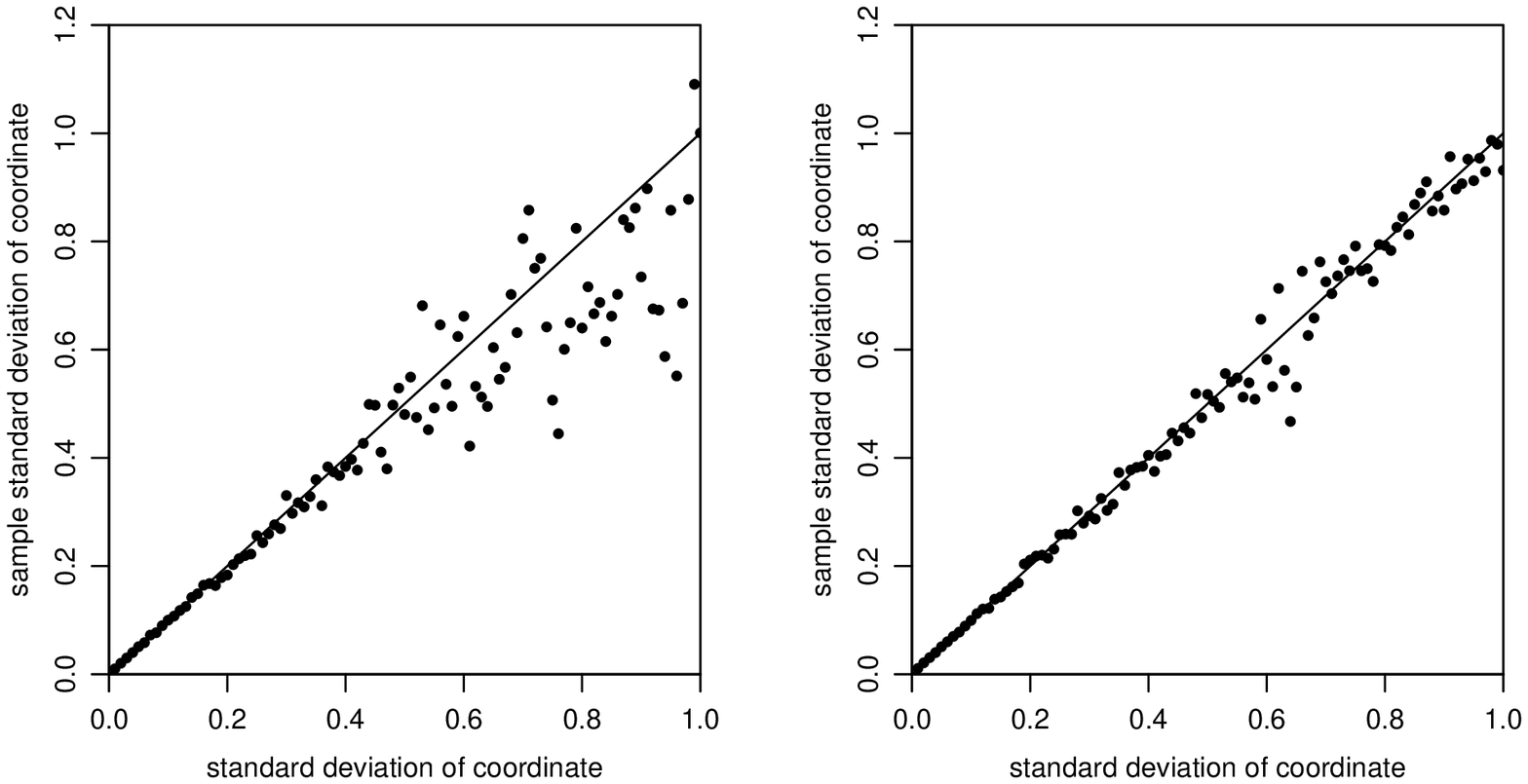}
  \vspace*{-20pt}
\end{center}
\caption{Estimates of means (top) and standard deviations (bottom) for the
100-dimensional example, using random-walk Metropolis (left) and HMC (right).
The 100 variables are labelled on the horizontal axes by the true standard
deviaton of that variable.  Estimates are on the vertical 
axes.}\label{fig:demo100d-est}
\end{figure}

Figure~\ref{fig:demo100d-trace} shows results from runs of 1000
iterations of HMC (right) and of random-walk Metropolis (left),
counting 150 random-walk Metropolis updates as one iteration, so that
the computation time per iteration is comparable to that for HMC.  The
plot shows the last variable, with the largest standard deviation.
The autocorrelation of these values is clearly much higher for
random-walk Metropolis than for HMC.  Figure \ref{fig:demo100d-est}
shows the estimates for the mean and standard deviation of each of the
100 variables obtained using the HMC and random-walk Metropolis runs
(estimates were just the sample means and sample standard deviations
of the values from the 1000 iterations).  Except for the first few
variables (with smallest standard deviations), the error in the mean
estimates from HMC is roughly 10 times less than the error in the mean
estimates from random-walk Metropolis.  The standard deviation
estimates from HMC are also better.

The randomization of the leapfrog stepsize done in this example
follows the advice discussed at the end of Section~\ref{ss:HMC}.  In
this example, not randomizing the stepsize (fixing
$\varepsilon=0.013$) does in fact cause problems --- the variables
with standard deviations near 0.31 or 0.62 change only slowly, since
$150$ leapfrog steps with $\epsilon=0.013$ produces nearly a full or
half cycle for these variables, so an accepted trajectory does not
make much of a change in the absolute value of the variable.

\section{HMC in practice and theory}\label{s:pr-th}

Obtaining the benefits from HMC illustrated in the previous section,
including random-walk avoidance, requires proper tuning of $L$ and
$\varepsilon$.  I discuss tuning of HMC below, and also show how
performance can be improved by using whatever knowledge is available
regarding the scales of variables and their correlations.  After
briefly discussing what to do when HMC alone is not enough, I discuss
an additional benefit of HMC --- its better scaling with
dimensionality than simple Metropolis methods.

\subsection{Effect of linear transformations}\label{ss:linear}

Like all MCMC methods I'm aware of, the performance of HMC may change
if the variables being sampled are transformed by multiplication by
some non-singular matrix, $A$.  However, performance stays the same
(except perhaps in terms of computation time per iteration) if at the
same time the corresponding momentum variables are multiplied by
$(A\T)^{-1}$.  These facts provide insight into the operation of HMC,
and can help us improve performance when we have some knowledge of the
scales and correlations of the variables.  

Let the new variables be $q' = Aq$. The probability density for $q'$
will be given by $P'(q') \,=\, P(A^{-1}q')\,/\,|\!\det(A)|$, where
$P(q)$ is the density for $q$.  If the distribution for $q$ is the
canonical distribution for a potential energy function $U(q)$ (see
Section~\ref{ss:prob}), we can obtain the distribution for $q'$ as the
canonical distribution for $U'(q') = U(A^{-1}q')$.  (Since $|\!\det(A)|$ is a
constant, we needn't include a $\log |\!\det(A)|$ term in the potential energy.)

We can choose whatever distribution we wish for the
corresponding momentum variables, so we could decide to use the same
kinetic energy as before.  Alternatively, we can choose to transform
the momentum variables by $p' = (A\T)^{-1}p$, and use a new kinetic energy
of $K'(p') = K(A\T p')$.  If we were using a quadratic kinetic energy,
$K(p)=p\T M^{-1} p\, /\, 2$ (see equation~\eqref{e:quadK}), the 
new kinetic energy will be
\beq
   K'(p') & = & (A\T p')\T M^{-1} (A\T p')\,/\,2 
          \ \ =\ \  (p')\T (A\, M^{-1} A\T)\, p'\,/\, 2
          \ \ =\ \  (p')\T (M')^{-1} p'\,/\, 2 \ \ \ \ \ \
   \label{e:Ktrans}
\eeq
where $M' = (A\, M^{-1} A\T)^{-1} = (A^{-1})\T M A^{-1}$.  

If we use momentum variables transformed in this way,
the dynamics for the new variables, $(q',p')$, essentially
replicates the original dynamics for $(q,p)$, so the performance of
HMC will be the same.  To see this, note that if we follow Hamiltonian
dynamics for $(q',p')$, the result in terms of the original variables
will be as follows (see equations~\eqref{e:ham-eq-b1} and~\eqref{e:ham-eq-b2}):
\beq
  {dq \over dt} & = & A^{-1}\, {dq' \over dt} 
    \ \ =\ \ A^{-1} (M')^{-1}\, p'
    \ \ =\ \ A^{-1} (A\,M^{-1} A\T) (A\T)^{-1}\, p
    \ \ =\ \ M^{-1}\, p\ \ \ \\[5pt]
  {dp \over dt} & = & A\T\, {dp' \over dt} 
    \ \ =\ \ - A\T\, \nabla U'(q')
    \ \ =\ \ - A\T\, (A^{-1})\T\,\nabla U(A^{-1}q')
    \ \ =\ \ - \nabla U(q) \ \ \ 
\eeq
which matches what would happen following Hamiltonian dynamics for $(q,p)$.

If $A$ is an orthogonal matrix (such as a rotation matrix), for which
$A^{-1} = A\T$, the performance of HMC is unchanged if we transform
both $q$ and $p$ by multiplying by $A$ (since $(A\T)^{-1}=A$). If we
chose a rotationally symmetric distribution for the momentum,
with $M=m I$ (ie, the momentum variables are independent, each having
variance $m$), such an orthogonal transformation will not change the
kinetic energy function (and hence not change the distribution of the
momentum variables), since we will have $M' = (A\,(mI)^{-1} A\T)^{-1}
= mI$.  

Such an invariance to rotation holds also for a random-walk Metropolis
method in which the proposal distribution is rotationally symmetric
(eg, Gaussian with covariance matrix $mI$).  In contrast, Gibbs
sampling is not rotationally invariant, nor is a scheme in which the
Metropolis algorithm is used to update each variable in turn (with
a proposal that changes only that variable).  However, Gibbs sampling is
invariant to rescaling of the variables (transformation by a diagonal
matrix), which is not true for HMC or random-walk Metropolis, unless
the kinetic energy or proposal distribution is transformed in a
corresponding way.  


Suppose we have an estimate, $\Sigma$, of the covariance matrix for
$q$, and suppose also that $q$ has at least a roughly Gaussian
distribution.  How can we use this information to improve the
performance of HMC?  One way is to transform the variables so that
their covariance matrix is close to the identity, by finding the
Cholesky decomposition, $\Sigma = LL\T$, with $L$ being
lower-triangular, and letting $q' = L^{-1} q$.  We then let our
kinetic energy function be $K(p)\, =\, p\T p\,/\,2$.  Since the
momentum variables are independent, and the position variables are
close to independent with variances close to one (if our estimate
$\Sigma$, and assumption that $q$ is close to Gaussian are good), HMC
should perform well using trajectories with a small number of leapfrog
steps, which will move all variables to a nearly independent point.
More realistically, the estimate $\Sigma$ may not be very good, but
this transformation could still improve performance compared to using
the same kinetic energy with the original $q$ variables.

An equivalent way to make use of the estimated covariance $\Sigma$ is
to keep the original $q$ variables, but use the kinetic energy
function $K(p) \,=\, p\T \Sigma p / 2$ --- ie, we let the momentum
variables have covariance $\Sigma^{-1}$.  The equivalence can be seen
by transforming this kinetic energy to correspond to a transformation
to $q' \,=\, L^{-1} q$ (see equation~\eqref{e:Ktrans}), which gives
$K(p') = (p')\T M'^{-1} p'$ with $M' \,=\, (L^{-1} (LL\T)
(L^{-1})\T)^{-1} \,=\, I$.  

Using such a kinetic energy function to compensate for correlations
between position variables has a long history in molecular dynamics
\citep{Bennett:1975}.  The usefulness of this technique is limited by
the computational cost of matrix operations when the dimensionality is
high.  

Using a diagonal $\Sigma$ can be feasible even in high-dimensional
problems.  Of course, this provides information only about the
different scales of the variables, not their correlation.  Moreover,
when the actual correlations are non-zero, it is not clear what scales
to use.  Making an optimal choice is probably infeasible.  Some
approximation to the conditional standard deviation of each variable
given all the others may be possible --- as I have done for Bayesian
neural network models \citep{neal:1996}.  If this also is not feasible,
using approximations to the marginal standard deviations of the
variables may be better than using the same scale for them all.

\subsection{Tuning HMC}\label{ss:tuning}

One practical impediment to the use of Hamiltonian Monte Carlo is the
need to select suitable values for the leapfrog stepsize,
$\varepsilon$, and the number of leapfrog steps, $L$, which together
determine the length of the trajectory in fictitious time,
$\varepsilon L$.  Most MCMC methods have parameters that need to be
tuned, with the notable exception of Gibbs sampling when the
conditional distributions are amenable to direct sampling.  However,
tuning HMC is more difficult in some respects than tuning a simple
Metropolis method.

\paragraph{Preliminary runs and trace plots.} 
Tuning HMC will usually require preliminary runs with trial values for
$\varepsilon$ and $L$.  In judging how well these runs work, trace
plots of quantities that are thought to be indicative of overall
convergence should be examined.  For Bayesian inference problems,
high-level hyperparameters are often among the slowest-moving
quantities.  The value of the potential energy function, $U(q)$, is
also usually of central significance.  The autocorrelation for such
quantities indicates how well the Markov chain is exploring the state
space.  Ideally, we would like the state after one HMC iteration to be
nearly independent of the previous state.

Unfortunately, preliminary runs can be misleading, if they are not long
enough to have reached equilibrium.  It is possible that the best
choices of $\varepsilon$ and $L$ for reaching equilibrium are
different from the best choices once equilibrium is reached, and even
at equilibrium, it is possible that the best choices vary from one
place to another.  If necessary, at each iteration of HMC,
$\varepsilon$ and $L$ can be chosen randomly from a selection of
values that are appropriate for different parts of the state space (or
these selections and can be used sequentially).

Doing several runs with different random starting states is advisable
(for both preliminary and final runs), so that problems with isolated
modes can be detected.  Note that HMC is no less (or more) vulnerable
to problems with isolated modes than other MCMC methods that make
local changes to the state.  If isolated modes are found to exist,
something needs to be done to solve this problem --- just combining
runs that are each confined to a single mode is not valid.  A
modification of HMC with ``tempering'' along a trajectory
(Section~\ref{ss:temper}) can sometimes help with multiple modes.

\paragraph{What stepsize?}

Selecting a suitable leapfrog stepsize, $\varepsilon$, is crucial.
Too large a stepsize will result in a very low acceptance rate for
states proposed by simulating trajectories.  Too small a stepsize will
either waste computation time, by the same factor as the stepsize is
too small, or (worse) will lead to slow exploration by a random walk,
if the trajectory length, $\varepsilon L$, is then too short (ie, $L$
is not large enough, see below).

Fortunately, as illustrated in Figure~\ref{fig:demo2d-a}, the choice
of stepsize is almost independent of how many leapfrog steps are done.
The error in the value of the Hamiltonian (which will determine the
rejection rate) usually does not increase with the number of leapfrog
steps, \textit{provided} that the stepsize is small enough that the
dynamics is stable.

The issue of stability can be seen in a simple one-dimensional
problem in which the following Hamiltonian is used:
\beq
   H(q,p) &=& q^2 \,/\, 2\sigma^2 \ +\ p^2 \,/\, 2
\eeq
The distribution for $q$ that this defines is Gaussian with standard 
deviation $\sigma$.
A leapfrog step for this system (as for any quadratic Hamiltonian)
will be a linear mapping from $(q(t),p(t))$ to
$(q(t+\varepsilon),p(t+\varepsilon))$.  Referring to equations~\eqref{e:leap1}
to~\eqref{e:leap3}, we see that this mapping can be represented by
a matrix multiplication as follows:
\beq
  \left[\begin{array}{c} 
     q(t+\varepsilon) \\[4pt] p(t+\varepsilon) 
  \end{array}\right]
  & = &
  \left[\begin{array}{cc} 
     1-\varepsilon^2/2\sigma^2 & \varepsilon \\[4pt]
     \ -\varepsilon/\sigma^2 + \varepsilon^3/4\sigma^4 \ 
     & \ 1-\varepsilon^2/2\sigma^2\ 
  \end{array}\right]
  \left[\begin{array}{c} q(t) \\[4pt] p(t) \end{array}\right]
\eeq
Whether iterating this mapping leads to a stable trajectory, or one that
diverges to infinity, depends on the magnitudes of the eigenvalues of the
above matrix, which are
\beq
    (1-\varepsilon^2/2\sigma^2) \ \pm \ (\varepsilon/\sigma)
       \sqrt{\varepsilon^2/4\sigma^2-1}
\eeq
When $\varepsilon/\sigma>2$, these eigenvalues are real, and at least
one will have absolute value greater than one.  Trajectories computed
using the leapfrog method with this $\varepsilon$ will therefore be unstable.
When $\varepsilon/\sigma<2$, the eigenvalues are complex, and both have
squared magnitude of
\beq
  (1-\varepsilon^2/2\sigma^2)^2 
    \ +\ (\varepsilon^2/\sigma^2)\,(1-\varepsilon^2/4\sigma^2)
  & = & 1
\eeq
Trajectories computed with $\varepsilon<2\sigma$ are therefore stable.

For multi-dimensional problems in which the kinetic energy used is
$K(p)=p\T p/2$ (as in the example above), the stability limit for
$\varepsilon$ will be determined (roughly) by the width of the
distribution in the most constrained direction --- for a Gaussian
distribution, this would the square root of the smallest eigenvalue
of the covariance matrix for $q$.  Stability for more general
quadratic Hamiltonians with $K(p)= p\T M^{-1} p / 2$ can be determined
by applying a linear transformation that makes $K(p')=(p')\T p'/2$, as
discussed above in Section~\ref{ss:linear}.

When a stepsize, $\varepsilon$, that produces unstable trajectories is
used, the value of $H$ grows exponentially with $L$, and consequently
the acceptance probability will be extremely small.  For
low-dimensional problems, using a value for $\varepsilon$ that is just
a bit below the stability limit is sufficient to produce a good
acceptance rate.  For high-dimensional problems, however, the stepsize
may need to be reduced further than this to keep the error in $H$ to a
level that produces a good acceptance probability.  This is discussed
further in Section~\ref{ss:scaling}.

Choosing too large a value of $\varepsilon$ can have very bad effects
on the performance of HMC.  In this respect, HMC is more sensitive to
tuning than random-walk Metropolis. A standard deviation for proposals
needs to be chosen for random-walk Metropolis, but performance
degrades smoothly as this choice is made too large, without the sharp
degradation seen with HMC when $\varepsilon$ exceeds the stability
limit.  (However, in high-dimensional problems, the degradation in
random-walk Metropolis with too large a proposal standard deviation
can also be quite sharp, so this distinction becomes less clear.)

This sharp degradation in performance of HMC when the stepsize is too
big would not be a serious issue if the stability limit were constant
--- the problem would be obvious from preliminary runs, and so could
be fixed.  The real danger is that the stability limit may differ for
several regions of the state space that all have substantial
probability.  If the preliminary runs are started in a region where
the stability limit is large, a choice of $\varepsilon$ a bit less
than this limit might appear to be appropriate.  However, if this
$\varepsilon$ is above the stability limit for some other region, the
runs may never visit this region, even though it has substantial
probability, producing a drastically wrong result.  To see why this
could happen, note that if the run ever does visit the region where
the chosen $\varepsilon$ would produce instability, it will stay there
for a very long time, since the acceptance probability with that
$\varepsilon$ will be very small.  Since the method nevertheless
leaves the correct distribution invariant, it follows that the run
only rarely moves to this region from a region where the chosen
$\varepsilon$ leads to stable trajectories.  One simple context where
this problem can arise is when sampling from a distribution with very
light tails (lighter than a Gaussian distribution), for which the log
of the density will fall faster than quadratically.  In the tails, the
gradient of the log density will be large, and a small stepsize will
be needed for stability.  See \citet{roberts:1996} for a discussion of
this in the context of the Langevin method (see
Section~\ref{ss:lang}).

This problem can be alleviated by choosing $\varepsilon$ randomly from
some distribution.  Even if the mean of this distribution is too
large, suitably small values for $\varepsilon$ may be chosen
occasionally.  (See Section~\ref{ss:HMC} for another reason to
randomly vary the stepsize.)  The random choice of $\varepsilon$
should be done once at the start of a trajectory, not for every
leapfrog step, since even if all the choices are below the stability
limit, random changes at each step lead to a random-walk in the error
for $H$, rather than the bounded error that is illustrated in
Figure~\ref{fig:demo2d-a}.

The ``short-cut'' procedures described in Section~\ref{ss:short} can
be seen as ways of saving computation time when a randomly chosen
stepsize in inappropriate.

\paragraph{What trajectory length?}  Choosing a suitable trajectory
length is crucial if HMC is to explore the state space systematically,
rather than by a random walk.  Many distributions are difficult to
sample from because they are tightly constrained in some directions,
but much less constrained in other directions.  Exploring the less
constrained directions is best done using trajectories that are long
enough to reach a point that is far from the current point in that
direction.  Trajectories can be too long, however, as is illustrated
in Figure~\ref{fig:demo2d-a}.  The trajectory shown on the left of
that figure is a bit too long, since it reverses direction and then
ends at a point that might have been reached with a trajectory about
half its length.  If the trajectory were a bit longer, the result
could be even worse, since the trajectory would not only take longer
to compute, but might also end near its starting point.

For more complex problems, one cannot expect to select a suitable
trajectory length by looking at plots like Figure~\ref{fig:demo2d-a}.
Finding the linear combination of variables that is least confined
will be difficult, and will be impossible when, as is typical, the
least confined ``direction'' is actually a non-linear curve or
surface.

Setting the trajectory length by trial and error therefore seems
necessary.  For a problem thought to be fairly difficult, a trajectory
with $L=100$ might be a suitable starting point.  If preliminary runs
(with a suitable $\varepsilon$, see above) shows that HMC reaches a
nearly independent point after only one iteration, a smaller value of
$L$ might be tried next.  (Unless these ``preliminary'' runs are
actually sufficient, in which case there is of course no need to do
more runs.)  If instead there is high autocorrelation in the run with
$L=100$, runs with $L=1000$ might be tried next.

As discussed at the ends of Sections~\ref{ss:HMC}
and~\ref{ss:HMC-ill}, randomly varying the length of the trajectory
(over a fairly small interval) may be desirable, to avoid choosing a
trajectory length that happens to produce a near-periodicity for some
variable or combination of variables.

\paragraph{Using multiple stepsizes.}

Using the results in Section~\ref{ss:linear}, we can exploit
information about the relative scales of variables to improve the
performance of HMC.  This can be done in two equivalent ways.  If
$s_i$ is a suitable scale for $q_i$, we could transform $q$, by
setting $q'_i\, =\, q_i/s_i$, or we could instead use a kinetic energy
function of $K(p) = p\T M^{-1} p$ with $M$ being a diagonal matrix
with diagonal elements $m_i\,=\,1/s_i^2$.

A third equivalent way to exploit this information, which is often the most
convenient, is to use different stepsizes for different pairs
of position and momentum variables.  To see how this works, consider a
leapfrog update (following equations~\ref{e:leap1} to~\ref{e:leap3}) with
$m_i=1/s_i^2$:
\beq
  p_i(t+\varepsilon/2) & = &
    p_i(t) \ -\ (\varepsilon/2)\, {\partial U \over \partial q_i} (q(t)) 
  \\[6pt]
  q_i(t+\varepsilon) & = & 
    q_i(t) \ +\ \varepsilon\, s_i^2\, p_i (t+\varepsilon/2)
  \\[6pt]
  p_i(t+\varepsilon) & = &
    p_i(t+\varepsilon/2) \ -\ (\varepsilon/2)\, {\partial U \over \partial q_i} 
    (q(t+\varepsilon))
\eeq
Define $(q^{(0)},p^{(0)})$ to be the state at the beginning of the 
leapfrog step (ie, $(q(t),p(t))$), 
define $(q^{(1)},p^{(1)})$ to be the final state (ie,
$(q(t+\varepsilon),p(t+\varepsilon))$), and define $p^{(1/2)}$ to be
half-way momentum (ie, $p(t+\varepsilon/2)$).  We can now rewrite the leapfrog 
step above as\vspace*{-6pt}
\beq
  p^{(1/2)}_i & = &
    p_i^{(0)} \ -\ (\varepsilon/2)\, {\partial U \over \partial q_i} (q^{(0})
  \\[6pt]
  q_i^{(1)} & = & 
    q_i^{(0)} \ +\ \varepsilon\, s_i^2\, p_i^{(1/2)}
  \\[6pt]
  p_i^{(1)} & = &
    p_i^{(1/2)} \ -\ (\varepsilon/2)\, {\partial U \over \partial q_i} (q^{(1)})
\eeq
If we now define rescaled momentum variables, $\tilde p_i = s_i p_i$, and
stepsizes $\varepsilon_i = s_i \varepsilon$, we can write the leapfrog
update as\vspace*{-6pt}
\beq
  \tilde p^{(1/2)}_i & = &
    \tilde p_i^{(0)} 
     \ -\ (\varepsilon_i/2)\, {\partial U \over \partial q_i} (q^{(0})
  \\[6pt]
  q_i^{(1)} & = & 
    q_i^{(0)} \ +\ \varepsilon_i\, \tilde p_i^{(1/2)}
  \\[6pt]
  \tilde p_i^{(1)} & = &
    \tilde p_i^{(1/2)} 
      \ -\ (\varepsilon_i/2)\, {\partial U \over \partial q_i} (q^{(1)})
\eeq
This is just like a leapfrog update with all $m_i=1$, but with
different stepsizes for different $(q_i,p_i)$ pairs.  Of course, the
successive values for $(q,\tilde p)$ can no longer be interpreted as
following Hamiltonian dynamics at consistent time points, but that is
of no consequence for the use of these trajectories in HMC.  Note that
when we sample for the momentum before each trajectory, each $\tilde
p_i$ is drawn independently from a Gaussian distribution with mean zero
and variance one, regardless of the value of $s_i$.

This multiple stepsize approach is often more convenient, especially
when the estimated scales, $s_i$, are not fixed, as discussed in
Section~\ref{ss:hier}, and the momentum is only partially refreshed
(Section~\ref{ss:partial}).

\subsection{Combining HMC with other MCMC updates}\label{ss:comb}

For some problems, MCMC using Hamiltonian Monte Carlo alone will be
impossible or undesirable.  Two situations where non-HMC updates will
be necessary are when some of the variables are discrete, and when 
the derivatives of the log probability density with respect to some of
the variables are expensive or impossible to compute.  HMC can then be
feasibly applied only to the other variables.  Another example is when
special MCMC updates have been devised that may help convergence in
ways that HMC does not --- eg, by moving between otherwise isolated
modes --- but which are not a complete replacement for HMC.  As discussed
in Section~\ref{ss:hier} below, Bayesian hierarchical models may also be best
handled with a combination of HMC and other methods such as Gibbs sampling.

In such circumstances, one or more HMC updates for all or a subset of
the variables can be alternated with one or more other updates that
leave the desired joint distribution of all variables invariant.  The
HMC updates can be viewed as either leaving this same joint
distribution invariant, or as leaving invariant the conditional
distribution of the variables that HMC changes, given the current
values of the variables that are fixed during the HMC update. These
are equivalent views, since the joint density can be factored as this
conditional density times the marginal density of the variables that
are fixed, which is just a constant from the point of view of a single
HMC update, and hence can be left out of the potential energy
function.

When both HMC and other updates are used, it may be best to use
shorter trajectories for HMC than would be used if only HMC were being
done.  This allows the other updates to be done more often, which
presumably helps sampling.  Finding the optimal tradeoff is likely to
be difficult, however. A variation on HMC that reduces the need for
such a tradeoff is described below in Section~\ref{ss:partial}.

\subsection{Scaling with dimensionality}\label{ss:scaling}

In Section~\ref{ss:HMC-ill}, one of the main benefits of HMC was
illustrated --- its ability to avoid the inefficient exploration of
the state space via a random walk.  This benefit is present (in at
least some degree) for most practical problems.  For problems in which
the dimensionality is moderate to high, another benefit of HMC over
simple random-walk Metropolis methods is a slower increase in the
computation time needed (for a given level of accuracy) as the
dimensionality increases.  (Note that here I will consider only
sampling performance after equilibrium is reached, not the time needed
to approach equilibrium from some initial state not typical of the
distribution, which is harder to analyse.)

\paragraph{Creating distributions of increasing dimensionality by
replication.} To talk about how performance scales with dimensionality
we need to assume something about how the distribution changes with
dimensionality, $d$.

I will assume that dimensionality increases by adding independent
replicas of variables --- ie, the potential energy function for
$q=(q_1,\ldots,q_d)$ has the form $U(q) \,=\, \Sigma\, u_i(q_i)$, for
functions $u_i$ drawn independently from some distribution.  Of
course, this is not what any real practical problem is like, but it
may be a reasonable model of the effect of increasing dimensionality
for some problems --- for instance, in statistical physics, distant
regions of large systems are often nearly independent.  Note that the
independence assumption itself is not crucial, since as discussed in
Section~\ref{ss:linear}, the performance of HMC (and of simple
random-walk Metropolis) does not change if independence is removed by
rotating the coordinate system, provided the kinetic energy function
(or random-walk proposal distribution) is rotationally symmetric.

For distributions of this form, in which the variables are
independent, Gibbs sampling will perform very well (assuming it is
feasible), producing an independent point after each scan of all
variables.  Applying Metropolis updates to each variable separately
will also work well, provided the time for a single-variable update
does not grow with $d$.  However, these methods are not invariant to
rotation, so this good performance may not generalize to the more
interesting distributions for which we hope to obtain insight with the
analysis below.

\paragraph{Scaling of HMC and random-walk Metropolis.}  Here, I discuss
informally how well HMC and random-walk Metropolis scale with
dimension, loosely following \citet[Section III]{creutz:1988}.

To begin, Cruetz notes that the following
relationship holds when any Metropolis-style algorithm is used to sample 
a density $P(x) \,=\, (1/Z) \exp(-E(x))$:
\beq
  1 & = & \E[P(x^*)/P(x)] 
  \ \ =\ \  \E[\exp(-(E(x^*)-E(x)))] \ \ =\ \ \E[\exp(-\Delta)]
  \label{e:ave-diff}
\eeq
where $x$ is the current state, assumed to be distributed according
to $P(x)$, $x^*$ is the proposed state, and $\Delta \,=\, E(x^*)-E(x)$.
Jensen's inequality then implies that the expectation of the energy
difference is non-negative:
\beq
  \E[\Delta] & \ge & 0
\eeq
The inequality will usually be strict.

When $U(q) \,=\, \Sigma\, u_i(q_i)$, and proposals are produced
independently for each $i$, we can apply these relationships
either to a single variable (or pair of variables) or to the entire
state.  For a single variable (or pair), I will write $\Delta_1$ for
$E(x^*)-E(x)$, with $x=q_i$ and $E(x)=u_i(q_i)$, or $x=(q_i,p_i)$ and
$E(x)=u_i(q_i)+p_i^2/2$.  For the entire state, I will write $\Delta_d$
for $E(x^*)-E(x)$, with $x=q$ and $E(x)=U(q)$, or $x=(q,p)$ and
$E(x)=U(q)+K(p)$).  For both random-walk Metropolis and HMC,
increasing dimension by replicating variables will lead to increasing
energy differences, since $\Delta_d$ is the sum of $\Delta_1$ for each
variable, each of which has positive mean.  This will lead to a
decrease in the acceptance probability --- equal to
$\min(1,\exp(-\Delta_d))$ --- unless the width of the proposal
distribution or the leapfrog stepsize is decreased to compensate.

More specifically, for random-walk Metropolis with proposals that
change each variable independently, the difference in potential energy
between a proposed state and the current state will be the sum of
independent differences for each variable.  If we fix the standard
deviation, $\varsigma$, for each proposed change, the mean and the
variance of this potential energy difference will both grow linearly
with $d$, which will lead to a progressively lower acceptance rate.
To maintain reasonable performance, $\varsigma$ will have to decrease
as $d$ increases.  Furthermore, the number of iterations needed to
reach a nearly independent point will be proportional to
$\varsigma^{-2}$, since exploration is via a random walk.
  
Similarly, when HMC is used to sample from a distribution in which the
components of $q$ are independent, using the kinetic energy $K(p)\,=\,
\Sigma\,p_i^2/2$, the different $(q_i,p_i)$ pairs do not interact
during the simulation of a trajectory --- each $(q_i,p_i)$ pair
follows Hamiltonian dynamics according to just the one term in the
potential energy involving $q_i$ and the one term in the kinetic
energy involving $p_i$.  There is therefore no need for the length in
fictitious time of a trajectory to increase with dimensionality.
However, acceptance of the end-point of the trajectory is based
on the error in $H$ due to the leapfrog approximation, which is the
sum of the errors pertaining to each $(q_i,p_i)$ pair.  For a fixed
stepsize, $\varepsilon$, and fixed trajectory length, $\varepsilon L$,
both the mean and the variance of the error in $H$ grow linearly with
$d$.  This will lead to a progressively lower acceptance rate as
dimensionality increases, if it is not counteracted by a decrease in
$\varepsilon$.  The number of leapfrog steps needed to reach an
independent point will be proportional to $\varepsilon^{-1}$.

To see which method scales better, we need to determine how rapidly
we must reduce $\varsigma$ and $\varepsilon$ as $d$ increases, in
order to maintain a reasonable acceptance rate.  As $d$ increases and 
$\varsigma$ or $\varepsilon$ go to zero, $\Delta_1$ will
go to zero as well.  Using a second-order approximation
of $\exp(-\Delta_1)$ as $1-\Delta_1+\Delta_1^2/2$, together with
equation~\eqref{e:ave-diff}, we find that
\beq
   \E[\Delta_1] & \approx & \E[\Delta_1^2]\,/\,2
   \label{e:delta-m-s}
\eeq
It follows from this that the variance of $\Delta_1$ is
twice the mean of $\Delta_1$ (when $\Delta_1$ is small), which implies
that the variance of $\Delta_d$ is twice the mean of $\Delta_d$ (even
when $\Delta_d$ is not small).  To
achieve a good acceptance rate, we must therefore keep the mean of 
$\Delta_d$ near one, since a large mean will not be saved by a similarly large 
standard deviation (which would produce fairly frequent acceptances as
$\Delta_d$ occasionally takes on a negative value).  

For random-walk Metropolis with a symmetric proposal distribution, we
can see how $\varsigma$ needs to scale by directly averaging
$\Delta_1$ for a proposal and its inverse.  Let the proposal for one
variable be $x^* = x + c$, and suppose that $c=a$ and $c=-a$ are
equally likely.  Approximating $U(x^*)$ to second order as $U(x)\, +\,
c U'(x) \,+\, c^2 U''(x)/2$, we find that the average of
$\Delta_1\,=\,U(x^*)-U(x)$ over $c=a$ and $c=-a$ is $a^2 U''(x)$.
Averaging this over the distribution of $a$, with standard deviation
$\varsigma$, and over the distribution of $x$, we see that
$\E[\Delta_1]$ is proportional to $\varsigma^2$.  It follows that
$\E[\Delta_d]$ is proportional to $d\varsigma^2$, so we can maintain a
reasonable acceptance rate by letting $\varsigma$ be proportional to
$d^{-1/2}$.  The number of iterations needed to reach a nearly
independent point will be proportional to $\varsigma^{-2}$, which will
be proportional to $d$.  The amount of computation time needed will
typically be proportional to $d^2$.

As discussed at the end of Section~\ref{ss:disc}, the error
in $H$ when using the leapfrog discretization to simulate a trajectory
of a fixed length is proportional to $\varepsilon^2$ (for sufficiently
small $\varepsilon$).  The error in $H$ for a single $(q_i,p_i)$ pair
is the same as $\Delta_1$, so we see that $\Delta_1^2$ is proportional
to $\varepsilon^4$.  Equation~\ref{e:delta-m-s} then implies that
$\E[\Delta_1]$ is also proportional to $\varepsilon^4$.  The average
total error in $H$ for all variables, $\E[\Delta_d]$, will be
proportional to $d\varepsilon^4$, and hence we must make $\varepsilon$
be proportional to $d^{-1/4}$ to maintain a reasonable acceptance
rate.  The number of leapfrog updates to reach a nearly independent
point will therefore grow as $d^{1/4}$, and the amount of computation
time will typically grow as $d^{5/4}$, which is much better than the
$d^2$ growth for random-walk Metropolis.

\paragraph{Optimal acceptance rates.}  By extending the analysis
above, we can determine what the acceptance rate of proposals is when
the optimal choice of $\varsigma$ or $\epsilon$ is used.  This is
helpful when tuning the algorithms --- provided, of course, that the
distribution sampled is high-dimensional, and has properties that are
adequately modeled by a distribution with replicated variables.

To find this acceptance rate, we first note that since Metropolis
methods satisfy detailed balance, the probability of an accepted
proposal with $\Delta_d$ negative must be equal to the probability of
an accepted proposal with $\Delta_d$ positive.  Since all proposals
with negative $\Delta_d$ are accepted, the acceptance rate is simply
twice the probability that a proposal has a negative $\Delta_d$.
For large $d$, the Central Limit Theorem implies that the distribution
of $\Delta_d$ is Gaussian, since it is a sum of $d$ independent
$\Delta_1$ values.  (This assumes that the variance of each $\Delta_1$
is finite.)  We saw above that the variance of $\Delta_d$ is twice its
mean, $\E[\Delta_d] = \mu$.  The acceptance probability can therefore be written
as follows \citep{gupta:1990}, for large $d$:
\beq
 P(\mbox{accept}) 
 & \!=\! & 2\,\Phi\big((0-\mu)\,\big/\sqrt{2\mu}\,\big)
 \  =\  2\,\Phi\big(\!-\!\sqrt{\mu/2}\,\big)
 \  =\  a(\mu)
\label{e:delta-accept}
\eeq
where $\Phi(z)$ is the cumulative distribution function for a Gaussian
variable with mean zero and variance one.

For random-walk Metropolis, the cost to obtain an independent point will 
be proportional to $1/(a\varsigma^2)$, where $a$ is the acceptance rate.
We saw above that $\mu = \E[\Delta_d]$ is proportional to $\varsigma^2$, so
the cost follows the proportionality
\beq
  C_{\mbox{\small rw}} & \propto & 1 \,/\, (a(\mu)\mu)
  \label{e:met-cost}
\eeq
Numerical calculation shows that this is minimized when $\mu=2.8$ and
$a(\mu)=0.23$.

For HMC, the cost to obtain an independent point will be proportional
to $1/(a\varepsilon)$, and as we saw above, $\mu$ is proportional to
$\varepsilon^4$.  From this we obtain
\beq
  C_{\mbox{\tiny HMC}} & \propto & 1 \,/\, (a(\mu)\mu^{1/4})
  \label{e:hmc-cost}
\eeq
Numerical calculation shows that the minimum is when $\mu=0.41$ and
$a(\mu)=0.65$.

The same optimal 23\% acceptance rate for random-walk Metropolis was
previously obtained using a more formal analysis by
\citet{roberts:1997}.  The optimal 65\% acceptance rate for HMC that I
derive above is consistent with previous empirical results on 
distributions following the model here \citep[Figure 2]{neal:1994},
and on real high-dimensional problems (\citealp{creutz:1988}, Figures 2 and 3;
\citealp{sexton:1992}, Table 1). 
\citet{kennedy:1991} obtained explicit and rigorous results for
HMC applied to multivariate Gaussian distributions.

\paragraph{Exploring the distribution of potential energy.}  The
better scaling behaviour of HMC seen above depends crucially on the
resampling of momentum variables.  We can see this by considering how
well the methods explore the distribution of the potential energy,
$U(q)\,=\, \Sigma\, u_i(q_i)$.  Because $U(q)$ is a sum of $d$ independent 
terms, its standard deviation will grow in proportion to $d^{1/2}$.  

Following \citet{caracciolo:1994}, we note that the expected change in
potential energy from a single Metropolis update will be no more than
order one --- intuitively, large upwards changes are unlikely to be
accepted, and since Metropolis updates satisfy detailed balance, large
downward changes must also be rare (in equilibrium).  Because changes
in $U$ will follow a random walk (due again to detailed balance), it
will take at least order $(d^{1/2}/\,1)^2 \,=\, d$ Metropolis updates
to explore the distribution of $U$. 

In the first step of an HMC iteration, the resampling of momentum
variables will typically change the kinetic energy by an amount that
is proportional to $d^{1/2}$, since the kinetic energy is also a sum
of $d$ independent terms, and hence has standard deviation that grows
as $d^{1/2}$ (more precisely, its standard deviation is
$d^{1/2}/2^{1/2}$).  If the second step of HMC proposes a distant point,
this change in kinetic energy (and hence in $H$) will tend, by the end
of the trajectory, to have become equally split between kinetic and
potential energy.  If the end-point of this trajectory is accepted,
the change in potential energy from a single HMC iteration will be
proportional to $d^{1/2}$, comparable to its overall range of
variation.  So, in contrast to random-walk Metropolis, we may hope
that only a few HMC iterations will be sufficient to move to a nearly
independent point, even for high-dimensional distributions.

Analysing how well methods explore the distribution of $U$ can also
provide insight into their performance on distributions that aren't
well modeled by replication of variables, as we will see in the next
section.

\subsection{HMC for hierarchical models}\label{ss:hier}

Many Bayesian models are defined hierarchically.  A large set of
low-level parameters have prior distributions that are determined by
fewer higher-level ``hyperparameters'', which in turn may have priors
determined by yet-higher-level hyperparameters.  For example, in a
regression model with many predictor variables, the regression
coefficients might be given Gaussian prior distributions, with mean
of zero and a variance that is a hyperparameter.  This
hyperparameter could be given a broad prior distribution, so that its
posterior distribution is determined mostly by the data.

One could apply HMC to these models in an obvious way (after taking
the logs of variance hyperparameters, so they will be unconstrained).
However, it may be better to apply HMC only to the lower-level
parameters, for reasons I will now discuss.  (See
Section~\ref{ss:comb} for general discussion of applying HMC to a
subset of variables.)

I will use my work on
Bayesian neural network models \citep{neal:1996} as an example.  Such
models typically have several groups of low-level parameters, each
with an associated variance hyperparameter.  The posterior
distribution of these hyperparameters reflects important aspects of
the data, such as which predictor variables are most relevant to the
task.  The efficiency with which values for these hyperparameters are
sampled from the posterior distribution can often determine the
overall efficiency of the MCMC method.

I use HMC only for the low-level parameters in Bayesian neural network
models, with the hyperparameters being fixed during an HMC update.
These HMC updates alternate with Gibbs sampling updates of the
hyperparameters, which (in the simpler versions of the models) are
independent given the low-level parameters, and have conditional
distributions of standard form.  By using HMC only for the low-level
parameters, the leapfrog stepsizes used can be set using heuristics
that are based on the current hyperparameter values.  (I use the
multiple stepsize approach described at the end of
Section~\ref{ss:tuning}, equivalent to using different mass values,
$m_i$, for different parameters.)  For example, the size of the
network ``weights'' on connections out of a ``hidden unit'' determine
how sensitive the likelihood function is to changes in weights on
connections into the hidden unit; the variance of the weights on these
outgoing connections is therefore useful in setting the stepsize for
the weights on the incoming connections.  If the hyperparameters were
changed by the same HMC updates as change the lower-level parameters,
using them to set stepsizes would not be valid, since a reversed
trajectory would use different stepsizes, and hence not retrace the
original trajectory.  Without a good way to set stepsizes, HMC for the
low-level parameters would likely be much less efficient.

\citet{choo:2000} bypassed this problem by using a modification of HMC
in which trajectories are simulated by alternating leapfrog steps that
update only the hyperparameters with leapfrog steps that update only
the low-level parameters.  This procedure maintains both reversibility
and volume-preservation (though not necessarily symplecticness), while
allowing the stepsizes for the low-level parameters to be set using
the current values of the hyperparameters (and vice versa).  However,
performance did not improve as hoped because of a second issue with
hierarchical models.

In these Bayesian neural network models, and many other hierarchical
models, the joint distribution of both low-level parameters and
hyperparameters is highly skewed, with the probability density varying
hugely from one region of high posterior probability to another.
Unless the hyperparameters controlling the variances of low-level
parameters have very narrow posterior distributions, the joint
posterior density for hyperparameters and low-level parameters will
vary greatly from when the variance is low to when it is high.  

For instance, suppose that in its region of high posterior
probability, a variance hyperparameter varies by a factor of four. If
this hyperparameter controls 1000 low-level parameters, their typical
prior probability density will vary by a factor of $2^{1000} =
1.07\times10^{301}$, corresponding to a potential energy range of
$\log(2^{1000}) = 693$, with a standard deviation of $693/12^{1/2} = 200$
(since the variance of a uniform distribution is one twelfth of its
range).  As discussed at the end of Section~\ref{ss:scaling}, one HMC
iteration changes the energy only through the resampling of the
momentum variables, which at best leads to a change in potential
energy with standard deviation of about $d^{1/2}/2^{3/2}$.  For this
example, with 1000 low-level parameters, this is 11.2, so about
$(200/11.2)^2 = 319$ HMC iterations will be needed to reach an
independent point.  

One might obtain similar performance for this example using Gibbs
sampling.  However, for neural network models, there is no feasible
way of using Gibbs sampling for the posterior distribution of the
low-level parameters, but HMC can be applied to the conditional
distribution of the low-level parameters given the hyperparameters.
Gibbs sampling can then be used to update the hyperparameters.  As we
have seen, performance would not be improved by trying to update the
hyperparameters with HMC as well, and updating them by Gibbs sampling
is easier.

\citet{choo:2000} tried another approach that could potentially
improve on this --- reparameterizing low-level parameters $\theta_i$,
all with variance $\exp(\kappa)$, by letting $\theta_i \,=\,
\phi_i\exp(\kappa/2)$, and then sampling for $\kappa$ and the $\phi_i$
using HMC.  The reparameterization eliminates the extreme variation in
probability density that HMC cannot efficiently sample.  However, he
found that it is difficult to set a suitable stepsize for $\kappa$,
and that the error in $H$ tended to grow with trajectory length,
unlike the typical situation when HMC is used only for the low-level
parameters.  Use of ``tempering'' techniques (see
Section~\ref{ss:temper}) is another possibility.

Even though it does not eliminate all difficulties, HMC is very useful
for Bayesian neural network models --- indeed, without it, they might
not be feasible for most applications.  Using HMC for at least the
low-level parameter can produce similar benefits for other
hierarchical models \citep[eg,][]{ishwaran:1999}, especially when the
posterior correlations of these low-level parameters are high.  As in
any application of HMC, however, careful tuning of the stepsize and
trajectory length is generally necessary.

\section{Extensions and variations on HMC}\label{s:var}

The basic HMC algorithm of Figure~\ref{f:HMC} can be modified in many
ways, either to improve its efficiency, or to make it useful for a wider
range of distributions.  In this section, I will start by discussing
alternatives to the leapfrog discretization of Hamilton's equations,
and also show how HMC can handle distributions with constraints on the
variables (eg, variables that must be positive).  I will then discuss
a special case of HMC --- when only one leapfrog step is done --- and
show how it can be extended to produce an alternative method of
avoiding random walks, which may be useful when not all variables are
updated by HMC.  Most applications of HMC can benefit from using a
variant in which ``windows'' of states are used to increase the
acceptance probability.  Another widely applicable technique is to use
approximations to the Hamiltonian to compute trajectories, while still
obtaining correct results by using the exact Hamiltonian when deciding
whether to accept the endpoint of the trajectory.  Tuning of HMC may
be assisted by using a ``short-cut'' method that avoids computing the
whole trajectory when the stepsize is inappropriate.  Tempering
methods have potential to help with distributions having multiple
modes, or which are highly skewed.

There are many other variations that I will not be able to review
here, such as the use of a ``shadow Hamiltonian'' that is exactly
conserved by the inexact simulation of the real Hamiltonian
\citep{izaguirre-hampton:2004}, and the use of symplectic integration
methods more sophisticated than the leapfrog method
\cite[eg,][]{creutz-gocksch:1989}, including a recent proposal by
\citet{girolami-etal:2009} of a symplectic integrator for a
non-separable Hamiltonian in which $M$ in the kinetic energy of
\eqref{e:quadK} depends on $q$, allowing for ``adaptation'' based on
local information.

\subsection{Discretization by splitting: handling constraints and
            other applications}\label{ss:split}

The leapfrog method is not the only discretization of Hamilton's
equations that is reversible and volume-preserving, and hence can be
used for Hamiltonian Monte Carlo.  Many ``symplectic integration
methods'' have been devised, mostly for applications other than HMC
(eg, simulating the solar system for millions of years to test its
stability).  It is possible to devise methods that have a higher order
of accuracy than the leapfrog method \citep[for example,
see][]{mclachlan:1992}.  Using such a method for HMC will produce
asymptotically better performance than the leapfrog method, as
dimensionality increases.  Experience has shown, however, that the
leapfrog method is hard to beat in practice.

Nevertheless, it is worth taking a more general look at how
Hamiltonian dynamics can be simulated, since this also points to how
constraints on the variables can be handled, as well as possible
improvements such as exploiting partial analytic solutions.

\paragraph{Splitting the Hamiltonian.}  Many symplectic
discretizations of Hamiltonian dynamics can be derived by
``splitting'' the Hamiltonian into several terms, and then for each
term in succession, simulating the dynamics defined by that term for
some small time step, then repeating this procedure until the desired
total simulation time is reached.  If the simulation for each term can
be done analytically, we obtain a symplectic approximation to the
dynamics that is feasible to implement.

This general scheme is described by \citet[Section 4.2]{leimkuhler:2004} and 
by \citet{sexton:1992}.  Suppose that the Hamiltonian can be written as
a sum of $k$ terms, as follows:
\beq
   H(q,p) & = & 
    H_1(q,p) \ +\ H_2(q,p) \ +\ \cdots\ +\ H_{k-1}(q,p) \ +\ H_k(q,p)
\eeq
Suppose also that we can \textit{exactly} implement Hamiltonian
dynamics based on each $H_i$, for $i=1,\ldots,k$, with
$T_{i,\varepsilon}$ being the mapping defined by applying dynamics
based on $H_i$ for time $\varepsilon$.  As shown by
\citeauthor{leimkuhler:2004}, if the $H_i$ are twice differentiable,
the composition of these mappings, $T_{1,\varepsilon} \,\circ\,
T_{2,\varepsilon} \,\circ\, \cdots \,\circ\, T_{k-1,\varepsilon}
\,\circ\, T_{k,\varepsilon}$, is a valid discretization of Hamiltonian
dynamics based on $H$, which will reproduce the exact dynamics in the
limit as $\varepsilon$ goes to zero, with global error of order $\varepsilon$
or less.

Furthermore, this discretization will preserve volume, and will be
symplectic, since these properties are satisfied by each of the
$T_{i,\varepsilon}$ mappings.  The discretization will also be
reversible if the sequence of $H_i$ is symmetrical --- ie, $H_i(q,p) =
H_{k-i+1}(q,p)$.  As mentioned at the end of Section~\ref{ss:disc},
any reversible method must have global error of even order in
$\varepsilon$ \citep[Section 4.3.3]{leimkuhler:2004}, which means the
global error must be of order $\varepsilon^2$ or better.

We can derive the leapfrog method from a symmetrical splitting of the
Hamiltonian.  If $H(q,p) \,=\, U(q) + K(p)$, we can write the
Hamiltonian as
\beq
   H(q,p) & = & U(q)/2 \ +\ K(p) \ +\  U(q)/2
\eeq
which corresponds to a split with $H_1(q,p) = H_3(q,p) = U(q)/2$ and 
$H_2(q,p) = K(p)$.  
Hamiltonian dynamics based on $H_1$ is (equations~\eqref{e:ham-eq-a1} 
and~\eqref{e:ham-eq-a2}):
\beq
  {dq_i \over dt} & = & {\partial H_1 \over \partial p_i} \ \ =\ \ 0 \\[4pt]
  {dp_i \over dt} & = & - {\partial H_1 \over \partial q_i} 
    \ \  =\ \ - {1 \over 2} {\partial U \over \partial q_i} 
\eeq
Applying this dynamics for time $\varepsilon$ just adds
$-(\varepsilon/2)\, \partial U/\partial q_i$ to each $p_i$,
which is the first part of a leapfrog step (equation~\eqref{e:leap1}).
The dynamics based on $H_2$ is as follows:
\beq
  {dq_i \over dt} & = & {\partial H_2 \over \partial p_i}
    \ \  =\ \ {\partial K \over \partial p_i} \\[4pt]
  {dp_i \over dt} & = & - {\partial H_2 \over \partial q_i}  \ \ =\ \ 0 
\eeq
If $K(p)=(1/2)\sum p_i^2/m_i$, applying this dynamics for time
$\epsilon$ results in adding $\varepsilon p_i/m_i$ to each $q_i$,
which is the second part of a leapfrog step
(equation~\eqref{e:leap2}).  Finally, $H_3$ produces the third part of
a leapfrog step (equation~\eqref{e:leap3}), which is the same as the
first part, since $H_3=H_1$.

\paragraph{Splitting to exploit partial analytical solutions.}

One situation where splitting can help is when the potential energy
contains a term that can, on its own, be handled analytically.  For
example, the potential energy for a Bayesian posterior distribution
will be the sum of minus the log prior density for the parameters and
minus the log likelihood.  If the prior is Gaussian, the log prior
density term will be quadratic, and can be handled analytically (eg,
see the one dimensional example at the end of Section~\ref{ss:eq}).

We can modify the leapfrog method for this situation by using a
modified split.  Suppose $U(q) \,=\, U_0(q) + U_1(q)$, with $U_0$
being analytically tractable, in conjunction with the kinetic energy,
$K(p)$.  We use the split
\beq
   H(q,p) & = & U_1(q)/2 \ +\ \big[U_0(q)+K(p)\big] \ +\  U_1(q)/2
\label{e:split-an}
\eeq
Ie, $H_1(q,p) = H_3(q,p) = U_1(q)/2$ and $H_2(q,p) = U_0(q)+K(p)$.
The first and last half-steps for $p$ are the same as for ordinary
leapfrog, based on $U_1$ alone.  The middle full step for $q$, which
in ordinary leapfrog just adds $\varepsilon p$ to $q$, is replaced
by the analytical solution for following the exact dynamics based on the 
Hamiltonian $U_0(q)+K(p)$ for time $\varepsilon$.

With this procedure, it should be possible to use a larger stepsize
(and hence use fewer steps in a trajectory), since part of the
potential energy has been separated out and handled exactly.  The
benefit of handling the prior exactly may be limited, however, since
the prior is usually dominated by the likelihood.  

\paragraph{Splitting potential energies with variable computational costs.}

Splitting can also help if the potential energy function can be
split into two terms, one of which requires less computation time to
evaluate than the other \citep{sexton:1992}.  
Suppose $U(q) \,=\, U_0(q) + U_1(q)$, with
$U_0$ being cheaper to compute than $U_1$, and let the kinetic energy
be $K(p)$.  We can use the following split, for some $M>1$:
\beq
   H(q,p) & = & U_1(q)/2 \ +\ 
                \sum_{m=1}^M \Big[ U_0(q)/2M \,+\, K(p)/M \,+\, U_0(q)/2M\Big] 
                \ +\  U_1(q)/2\ \ \
\eeq
We label the $k=3M+2$ terms as $H_1(q,p) = H_k(q,p) = U_1(q)/2$ and for 
$i=1,\ldots,M$, $H_{3i-1}(q,p) = H_{3i+1}(q,p) = U_0(q)/2M$ and
$H_{3i}(q,p) = K(p)/M$.  The resulting discretization can be seen as
a nested leapfrog method.  The $M$ inner leapfrog steps involve only $U_0$,
and use an effective stepsize of $\varepsilon/M$. The outer leapfrog step
takes half steps for $p$ using only $U_1$, and replaces the update for $q$ in 
the middle with the $M$ inner leapfrog steps.  

If $U_0$ is much cheaper to compute than $U_1$, we can use a large
value for $M$ without increasing computation time by much.  The
stepsize, $\varepsilon$, that we can use will then be limited mostly
by the properties of $U_1$, since the effective stepsize for $U_0$ is
much smaller, $\varepsilon/M$.  Using a bigger $\varepsilon$ than with
the standard leapfrog method will usually be possible, and hence we
will need fewer steps in a trajectory, with fewer computations of
$U_1$.

\paragraph{Splitting according to data subsets.}

When sampling from the posterior distribution for a Bayesian model of
independent data points, it may be possible to save computation time
by splitting the potential energy into terms for subsets of the data.

Suppose we partition the data into subsets $S_m$, for $i=1,\ldots,M$,
typically of roughly equal size.
We can then write the log likelihood function as $\ell(q) \,=\,
\sum_{m=1}^M \ell_m(q)$, where $\ell_m$ is the log likelihood function
based on the data points in $S_m$.  If $\pi(q)$ is the prior density
for the parameters, we can let $U_m(q)\, =\, -\log(\pi(q))/M - \ell_m(q)$,
and split the Hamiltonian as follows:\vspace*{-6pt}
\beq
  H(q,p) & = & \sum_{m=1}^M \Big[ U_m(q)/2 + K(p)/M + U_m(q)/2 \Big]
\eeq
Ie, we let the $k=3M$ terms be $H_{3m-2}(q,p) = H_{3m}(q,p) =
U_m(q)/2$ and $H_{3m-1}(q,p)=K(p)/m$.  The resulting discretization
with stepsize $\varepsilon$ effectively performs $M$ leapfrog steps
with stepsize $\varepsilon/M$, with the $m$th step using $MU_m$ as the
potential energy function.

This scheme can be beneficial if the data set is redundant, with many
data points that are similar.  We then expect $MU_m(q)$ to be
approximately the same as $U(q)$, and we might hope that we could set
$\varepsilon$ to be $M$ times larger than with the standard leapfrog
method, obtaining similar results with $M$ times less computation.  In
practice, however, the error in $H$ at the end of the trajectory will
be larger than with standard leapfrog, so the gain will be less than
this.  I found \citep[Sections 3.5.1 and 3.5.2]{neal:1996} that the
method can be beneficial for neural network models, especially when
combined with the windowed HMC procedure described below in
Section~\ref{ss:windows}.

Note that unlike the other examples above, this split is \textit{not}
symmetrical, and hence the resulting discretization is not reversible.
However, it can still be used to produce a proposal for HMC as long as
the labelling of the subsets is randomized for each iteration, so that
the reverse trajectory has the same probability of being produced as
the forward trajectory.  (It is possible, however, that some symmetrical
variation on this split might produce better results.)

\paragraph{Handling constraints.}

An argument based on splitting shows how to handle constraints on the
variables being sampled.  Here, I will consider only separate
constraints on some subset of the variables, with the constraint on
$q_i$ taking the form $q_i \le u_i$, or $q_i \ge l_i$, or both.  A
similar scheme can handle constraints taking the form $G(q) \ge 0$,
for any differentiable function $G$.

We can impose constraints on variables by letting the potential
energy be infinite for values of $q$ that violate any of the
constraints, which will give such points probability zero.  To see how
to handle such infinite potential energies, we can look at a limit of
potential energy functions that approach infinity, and the
corresponding limit of the dynamics.

To illustrate, suppose that $U_*(q)$ is the potential energy ignoring
constraints, and that $q_i$ is constrained to be less than $u_i$.
We can take the limit as $r\rightarrow\infty$ of the following
potential energy function (which is one of many that could be used):
\beq
  U(q) & = & U_*(q) \ +\ C_r(q_i,u_i), 
  \ \ \ \mbox{where} \ \ C_r(q_i,u_i) \ = \ \left\{\begin{array}{ll}
     0 & \mbox{if $q_i \le u_i$} \\[4pt]
     r^{r+1} (q_i-u_i)^r & \mbox{if $q_i > u_i$}
  \end{array}\right.\ \ \ \ \ \ \
\eeq
It is easy to see that $\lim_{r\rightarrow\infty} C_r(q_i,u_i)$ is zero for
any $q_i \le u_i$ and infinity for any $q_i > u_i$.  For any finite $r>1$, 
$U(q)$ is differentiable, so we can use it to define Hamiltonian dynamics.

To simulate the dynamics based on this $U(q)$, with a kinetic
energy $K(p)= (1/2) \sum p_i^2/m_i$, we can use the split of 
equation~\eqref{e:split-an}, with $U_1(q)=U_*(q)$ and $U_0(q)=C_r(q_i,u_i)$:
\beq
   H(q,p) & = & U_*(q)/2 \ +\ \big[C_r(q_i,u_i)+K(p)\big] \ +\  U_*(q)/2
\eeq
This produces a variation on the leapfrog method in which the
half-steps for $p$ (equations~\eqref{e:leap1} and~\ref{e:leap2})
remain the same, but the full step for $q$ (equation~\eqref{e:leap2})
is modified to account for the constraint on $q_i$.  After computing
$q_i' \,=\, q_i(t) + \varepsilon p_i(t+\varepsilon/2)/m_i$, we check if
$q_i'>u_i$.  If not, the value of $C_r(q_i,u_i)$ must be zero all
along the path from $q_i$ to $q_i'$, and we can set $q(t+\varepsilon)$ 
to $q_i'$.  But if $q_i'>u_i$, the dynamics based on the Hamiltonian
$C_r(q_i,u_i)+K(p)$ will be affected by the $C_r$ term.  This term
can be seen as a steep hill, which will be climbed as $q_i$ moves
past $u_i$, until the point is reached where $C_r$ is equal to
the previous value of $(1/2)p_i^2/m_i$, at which point $p_i$ will
be zero.  (If $r$ is sufficiently large, as it will be in the limit
as $r\rightarrow\infty$, this point will be reached
before the end of the full step.)  We will then fall down the hill, 
with $p_i$ taking on increasingly negative values, until we again reach 
$q_i=u_i$, when $p_i$ will be just the negative of the original value of 
$p_i$.  We then continue, now moving in the opposite direction, away from
the upper limit.

If several variables have constraints, we must follow this procedure
for each, and if a variable has both upper and lower constraints, we
must repeat the procedure until neither constraint is violated.  The
end result is that the full step for $q$ of equation~\eqref{e:leap2}
is replaced by the procedure shown in Figure~\ref{fig:constraints}.
Intuitively, the trajectory just bounces off the ``walls'' given by
the constraints.  If $U_*(q)$ is constant, these bounces are the
only interesting aspect of the dynamics, and the procedure is sometimes
referred to as ``billiards'' \citep[see, for example,][]{rujan:1997}.

\begin{figure}[t]
\begin{center}
\parbox{4.75in}{\texttt{
For each variable, $i=1,\ldots,d$:
\begin{itemize}
\item[1)] Let $p_i' \,=\, p_i(t+\varepsilon/2)$
\item[2)] Let $q_i' \,=\, q_i(t) \,+\, \varepsilon p_i'/m_i$
\item[3)] If $q_i$ is constrained, repeat the following until $q_i'$
          satisfies all constraints:
      \begin{itemize}
      \item[a)] If $q_i$ has an upper constraint, and $q_i'>u_i$\\[6pt]
                \mbox{~~} Let $q_i' \,=\, u_i \,-\, (q_i'-u_i)$
                and $p_i' \,=\, -p_i'$\vspace*{3pt}
      \item[b)] If $q_i$ has a lower constraint, and $q_i'<l_i$\\[6pt]
                \mbox{~~} Let $q_i' \,=\, l_i \,+\, (l_i-q_i')$ 
                and $p_i' \,=\, -p_i'$ 
      \end{itemize}
\item[4)] Let $q_i(t+\varepsilon)\,=\,q_i'$ and $p_i(t+\varepsilon/2)\,=\,p_i'$
\end{itemize}
}}\end{center}\vspace*{-8pt}
\caption{Modification to the leapfrog update of $q$ (equation~\eqref{e:leap2})
to handle constraints of the form $q_i \le u_i$ or $q_i \le l_i$.
}\label{fig:constraints}
\end{figure}

\subsection{Taking one step at a time --- the Langevin method}\label{ss:lang}

A special case of Hamiltonian Monte Carlo arises when the trajectory used
to propose a new state consists of only a single leapfrog step.  Suppose
that we use the kinetic energy $K(p)= (1/2) \sum p_i^2$.  An iteration
of HMC with one leapfrog step can be expressed in the following way.  We 
sample values for the momentum variables,
$p$, from their Gaussian distributions with mean zero and variance one, 
and then propose new values, $q^*$ and $p^*$, as follows:
\beq
   q_i^* & = & q_i 
      \ -\ {\varepsilon^{2\!\!} \over 2}\, {\partial U \over \partial q_i} (q)
      \ +\ \varepsilon p_i \label{e:lang} \\[6pt]
   p_i^* & = & p_i 
      \ -\ {\varepsilon \over 2}\, {\partial U \over \partial q_i} (q)
      \ -\ {\varepsilon \over 2}\, {\partial U \over \partial q_i} (q^*)
   \label{e:langp}
\eeq
We accept $q^*$ as the new state with probability 
\beq 
  \min\Big [1,\ 
    \exp\Big(\!-(U(q^*)-U(q))\,-\,{1\over2}\sum_i ((p_i^*)^2-p_i^2)\Big)\Big]
  \label{e:lang-accept}
  \\[-15pt]\nonumber
\eeq
and otherwise keep $q$ as the new state.

Equation~\eqref{e:lang} is known in physics as one type of ``Langevin
equation'', and this method is therefore known as \textit{Langevin
Monte Carlo (LMC)} in the the lattice field theory literature
\citep[eg,][]{kennedy:1990}.  

One can also remove any explicit mention of momentum variables, and
view this method as performing a Metropolis-Hastings update in which
$q^*$ is proposed from a Gaussian distribution where the $q_i^*$
are independent, with means of $q_i-(\varepsilon^2/2)[\partial U/\partial
q_i](q)$, and variances of $\varepsilon^2$.  Since this proposal is not
symmetrical, it must be accepted or rejected based on both the ratio
of the probability densities of $q^*$ and $q$ and on the ratio of
the probability densities for proposing $q$ from $q^*$ and vice versa
\citep{hastings:1970}.  To see the equivalence with HMC using one
leapfrog step, we can write the Metropolis-Hastings acceptance probability 
as follows:
\beq
  \min\left[1,\ 
     {\exp(-U(q^*)) \over \exp(-U(q))}\
       \prod_{i=1}^d \,
     {\exp\big(-\big(
         q_i-q_i^*+(\varepsilon^2/2)\,[\partial U/\partial q_i](q^*)\big)^2
       /\,2\varepsilon^2\big)
       \over
      \exp\big(-\big(
         q_i^*-q_i+(\varepsilon^2/2)\,[\partial U/\partial q_i](q)\big)^2
       /\,2\varepsilon^2\big)}
  \right]
  \label{e:lang-accept2}
\eeq
To see that this is the same as~\eqref{e:lang-accept}, note that
using equations~\eqref{e:lang} and~\eqref{e:langp}, we can write\vspace*{-6pt}
\beq
   p & = & {1\over\varepsilon}\left[q_i^*\,-\,q_i\,+\,{\varepsilon^2\over2}
             {\partial U \over \partial q_i}(q) \right] \\[6pt]
   p^* & = & - {1\over\varepsilon}\left[q_i\,-\,q_i^*\,+\,{\varepsilon^2\over2}
             {\partial U \over \partial q_i}(q^*) \right] 
\eeq
After substituting these into~\eqref{e:lang-accept}, it is straightforward
to see the equivalence to~\eqref{e:lang-accept2}.

In this Metropolis-Hastings form, the Langevin Monte Carlo method was
first proposed by Rossky, Doll, and Friedman (\citeyear{rossky:1978}),
for use in physical simulations.  Approximate Langevin methods without
an accept/reject step can also be used \citep[for a discussion of
this, see][Section 5.3]{neal:1993} --- as, for instance, in a paper on
statistical inference for complex models by \citet{grenander:1994},
where also an accept/reject step is proposed in the discussion by
J.~Besag (p.~591).

Although LMC can be seen as a special case of HMC, its properties are
quite different.  Since LMC updates are reversible, and generally make
only small changes to the state (since $\varepsilon$ typically cannot
be very large), LMC will explore the distribution via an inefficient
random walk, just like random-walk Metropolis updates.  

However, LMC has better scaling behaviour than random-walk Metropolis
as dimensionality increases, as can be seen from an analysis
paralleling that in Section~\ref{ss:scaling}
\citep{creutz:1988,kennedy:1990}.  The local error of the leapfrog
step is of order $\varepsilon^3$, so $\E[\Delta_1^2]$, the average
squared error in $H$ from one variable, will be of order
$\varepsilon^6$.  From equation~\eqref{e:delta-m-s}, $\E[\Delta]$ will
also be of order $\varepsilon^6$, and with $d$ independent variables,
$\E[\Delta_d]$ will be of order $d \varepsilon^6$, so that
$\varepsilon$ must scale as $d^{-1/6}$ in order to maintain a
reasonable acceptance rate.  Since LMC explores the distribution via a
random walk, the number of iterations needed to reach a nearly
independent point will be proportional to $\varepsilon^{-2}$, which
grows as $d^{1/3}$, and the computation time to reach a nearly
independent point grows as $d^{4/3}$.  This is better than the $d^2$
growth in computation time for random walk Metropolis, but worse than
the $d^{5/4}$ growth when HMC is used with trajectories that are long
enough to reach a nearly independent point.

We can also find what the acceptance rate for LMC will be when the
optimal $\varepsilon$ is used, when sampling a distribution with
independent variables replicated $d$ times.  As for random walk
Metropolis and HMC, the acceptance rate is given in terms of
$\mu=\E[\Delta_d]$ by equation~\eqref{e:delta-accept}.  The cost
of obtaining a nearly independent point using LMC is proportional
to $1/(a(\mu)\varepsilon^2)$, and since $\mu$ is proportional to
$\varepsilon^6$, we can write the cost as
\beq
  C_{\mbox{\tiny LMC}} & \propto & 1 \,/\, (a(\mu)\mu^{1/3})
  \label{e:lmc-cost}
\eeq
Numerical calculation shows that this is minimized when $a(\mu)$ is 0.57
a result obtained more formally by \citet{roberts:1998}.  This may be
useful for tuning, if the behaviour of LMC for the distribution being sampled 
resembles its behaviour when sampling for replicated independent variables.

\subsection{Partial momentum refreshment --- another way to avoid random walks}
\label{ss:partial}

The single leapfrog step used in the Langevin Monte Carlo algorithm
will usually not be sufficient to move to a nearly independent point,
so LMC will explore the distribution via an inefficient random walk.
This is why HMC is typically used with trajectories of many
leapfrog steps.  An alternative that can suppress random walk
behaviour even when trajectories consist of just one leapfrog step is
to only partially refresh the momentum between trajectories, as
proposed by \citet{horowitz:1991}.

Suppose that the kinetic energy has the typical form 
$K(p) \,=\, p\T M^{-1} p/2$.
The following update for $p$ will leave invariant
the distribution for the momentum
(Gaussian with mean zero and covariance $M$):\vspace*{-6pt}
\beq 
   p' & = & \alpha p \ +\ (1-\alpha^2)^{1/2} n 
\label{e:part-up}
\eeq
Here, $\alpha$ is any constant in the interval $[-1,+1]$, and $n$ is a
Gaussian random vector with mean zero and covariance matrix $M$.  To see
this, note that if $p$ has the required Gaussian distribution, the
distribution of $p'$ will also be Gaussian (since it is a linear
combination of independent Gaussians), will have mean $0$, and will
have covariance $\alpha^2 M + (1-\alpha^2) M\, =\, M$.

If $\alpha$ is only slightly less than one, $p'$ will be similar to
$p$, but repeated updates of this sort will eventually produce a value
for the momentum variables almost independent of the initial value.
When $\alpha=0$, $p'$ is just set to a random value drawn from its
Gaussian distribution, independent of its previous value.  Note that
when $M$ is diagonal, the update of each momentum variable, $p_i$, is
independent of the updates of other momentum variables.

The partial momentum update of equation~\eqref{e:part-up} can be substituted
for the full replacement of the momentum in the standard HMC algorithm.  This
gives a generalized HMC algorithm in which an iteration consists of three 
steps:\vspace{-8pt}
\begin{enumerate}
\item[1)] Update the momentum variables using equation~\eqref{e:part-up}.
          Let the new momentum be $p'$.
\item[2)] Propose a new state, $(q^*,p^*)$,
          by applying $L$ leapfrog steps with stepsize $\varepsilon$, starting 
          at $(q,p')$, and then
          negating the momentum.  Accept $(q^*,p^*)$ with probability
          \beq
            \min\Big[ 1,\, \exp(-U(q^*)+U(q)-K(p^*)+K(p')) \Big]
          \eeq
          If $(q^*,p^*)$ is accepted, let $(q'',p'') = (q^*,p^*)$; otherwise,
          let $(q'',p'') = (q,p')$.
\item[3)] Negate the momentum, so that the new state is 
          $(q'',-p'')$.\vspace{-8pt}
\end{enumerate}
The transitions in each of these steps ---
$(q,p)\rightarrow(q,p')$, $(q,p')\rightarrow (q'',p'')$, and 
$(q'',p'') \rightarrow (q'',-p'')$ --- leave the canonical distribution
for $(q,p)$ invariant.  The entire update therefore also leaves the
canonical distribution invariant.  The three transitions also 
each satisfy detailed balance, but the sequential combination of the
three does \textit{not} satisfy detailed balance (except when $\alpha=0$).
This is crucial, since if the combination were reversible, it would
still result in random walk behaviour when $L$ is small.  

Note that omitting step~(3) above would result in a valid algorithm,
but then, far from suppressing random walks, the method (with $\alpha$
close to one) would produce nearly back-and-forth motion, since the
direction of motion would reverse with every trajectory accepted in
step~(2).  With the reversal in step~(3), motion continues in the same
direction as long as the trajectories in step~(2) are accepted, since
the two negations of $p$ will cancel.  Motion reverses whenever a
trajectory is rejected, so if random walk behaviour is to be
suppressed, the rejection rate must be kept small.

If $\alpha=0$, the above algorithm is the same as standard HMC, since
step~(1) will completely replace the momentum variables, step~(2) is
the same as for standard HMC, and step~(3) will have no effect, since
the momentum will be immediately replaced anyway, in step (1)~of the
next iteration.

Since this algorithm can be seen as a generalization of standard HMC,
with an additional $\alpha$ parameter, one might think it will offer
an improvement, provided that $\alpha$ is tuned for best
performance. However, \citet{kennedy:2001} show that when the method
is applied to high-dimensional multivariate Gaussian distributions
only a small constant factor improvement is obtained, with no better
scaling with dimensionality.  Best performance is obtained using long
trajectories ($L$ large), and a value for $\alpha$ that is not very
close to one (but not zero, so the optimum choice is not standard
HMC).  If $L$ is small, the need to keep the rejection rate very low
(by using a small $\varepsilon$), as needed to suppress random walks,
makes the method less advantageous than standard HMC.

It is disappointing that only a small improvement is obtained with
this generalization when sampling a multivariate Gaussian, due to
limitations that likely apply to other distributions as well.
However, the method may be more useful than one would think from this.
For reasons discussed in Sections~\ref{ss:comb} and~\ref{ss:hier}, we
will often combine HMC updates with other MCMC updates (perhaps for
variables not changed by HMC).  There may then be a tradeoff between
using long trajectories to make HMC more efficient, and using shorter
trajectories so that the other MCMC updates can be done more often.
If shorter-than-optimal trajectories are to be used for this reason,
setting $\alpha$ greater than zero can reduce the random walk
behaviour that would otherwise result. 

Furthermore, rejection rates can be reduced using the ``window''
method described in the next section.  An analysis of partial momentum
refreshment combined with the window method might find that using
trajectories of moderate length in conjunction with a value for
$\alpha$ greater than zero produces a more substantial improvement.

\subsection{Acceptance using windows of states}\label{ss:windows}

Figure~\ref{fig:demo2d-a} (right plot) shows how the error in $H$
varies along a typical trajectory computed with the leapfrog method.
Rapid oscillations occur, here with a period of between 2 and 3
leapfrog steps, due to errors in simulating the motion in the most
confined direction (or directions, for higher-dimensional
distributions).  When a long trajectory is used to propose a state for
HMC, it is essentially random whether the trajectory ends at a state
where the error in $H$ is negative or close to zero, and hence will be
accepted with probability close to one, or whether it happens to end
at a state with a large positive error in $H$, and a correspondingly
lower acceptance probability.  If somehow we could smooth out these
oscillations, we might obtain a high probability of acceptance for
all trajectories.

I introduced a method for achieving this result that uses ``windows'' of
states at the beginning and end of the trajectory \citep{neal:1994}.
Here, I will present the method as an application of a general
technique in which we probabilistically map to a state in a different
space, perform a Markov chain transition in this new space, and then
probabilistically map back to our original state space \citep{neal:2006}.

Our original state space consists of pairs, $(q,p)$, of position and
momentum variables.  We will map to a sequence of $W$ pairs,
$[(q_0,p_0),\ldots,(q_{W-1},p_{W-1})]$, in which each $(q_i,p_i)$ for
$i \!>\! 0$ is the result of applying one leapfrog step (with some fixed
stepsize, $\varepsilon$) to $(q_{i-1},p_{i-1})$.  Note that even
though a point in the new space seems to consist of $W$ times as many
numbers as a point in the original space, the real dimensionality of
the new space is the same as the old, since the whole sequence of $W$
pairs is determined by $(q_0,p_0)$.

To probabilistically map from $(q,p)$ to a sequence of pairs,
$[(q_0,p_0),\ldots,(q_{W-1},p_{W-1})]$, we select $s$ uniformly from
$\{0,\ldots,W-1\}$, and set $(q_s,p_s)$ in the new state to our current
state $(q,p)$.  The other $(q_i,p_i)$ pairs in the new state are
obtained using leapfrog steps from $(q_s,p_s)$, for $i\!>\!s$, or
backwards leapfrog steps (ie, done with stepsize $-\varepsilon$) for
$i\!<\!s$.  It is easy to see, using the fact that leapfrog steps preserve
volume, that if our original state is distributed with probability
density $P(q,p)$, then the probability density of obtaining the
sequence $[(q_0,p_0),\ldots,(q_{W-1},p_{W-1})]$ by this procedure
is\vspace*{-6pt} 
\beq
  P([(q_0,p_0),\ldots,(q_{W-1},p_{W-1})]) & = & 
   {1 \over W} \sum_{i=0}^{W-1} P(q_i,p_i)
\label{e:seq-pd}
\eeq
since we can obtain this sequence from a $(q,p)$ pair that matches
any pair in the sequence, and the probability is $1/W$ that we will produce the 
sequence starting from each of these pairs (which happens 
only if the random selection of $s$ puts the pair at the right place in
the sequence).

Having mapped to a sequence of $W$ pairs, we now perform 
a Metropolis update that keeps the sequence distribution
defined by equation~\eqref{e:seq-pd} invariant, before mapping back to
the original state space.  To obtain a Metropolis proposal, we
perform $L-W+1$ leapfrog steps (for some $L \ge W\!-\!1$), starting from 
$(q_{W-1},p_{W-1})$, 
producing pairs $(q_W,p_W)$ to $(q_L,p_L)$.  We then
propose the sequence $[(q_L,-p_L),\ldots,(q_{L-W+1},-p_{L-W+1})]$.
We accept or reject this proposed sequence by the usual Metropolis criterion, 
with the acceptance probability being
\beq
  \min\left[ 1,\ 
   { \textstyle \sum_{i=L-W+1}^L P(q_i,p_i)_{\rule{0pt}{10pt}} \over
                \sum_{i=0}^{W-1\rule{0pt}{8pt}} P(q_i,p_i)} \,\right]
   \label{e:winacc}
\eeq
with $P(q,p) \propto \exp(-H(q,p))$.  (Note here that $H(q,p)=H(q,-p)$,
and that starting from the proposed sequence would lead symmetrically
to the original sequence being proposed.)

This Metropolis update leaves us with either the sequence
$[(q_L,p_L),\ldots,(q_{L-W+1},p_{L-W+1})]$ --- called the ``accept window'',
--- or the sequence $[(q_0,p_0),\ldots,(q_{W-1},p_{W-1})]$ --- called the
``reject window''.  (Note that these windows will overlap if
$L+1<2W$.)  We label the pairs in the window chosen as
$[(q^+_0,p^+_0),\ldots,(q^+_{W-1},p^+_{W-1})]$. We now produce a final
state for the windowed HMC update by probabilistically mapping from
this sequence to a single pair, choosing $(q^+_e,p^+_e)$
with probability
\beq
   { P(q^+_e,p^+_e) 
   \over \textstyle \sum_{i=0}^{W-1\rule{0pt}{8pt}} P(q^+_i,p^+_i) }
   \label{e:map-back}
\eeq
If the sequence in the chosen window was distributed according to
equation~\eqref{e:seq-pd}, the pair $(q^+_e,p^+_e)$ chosen 
will be distributed according to $P(q,p)
\propto \exp(-H(q,p))$, as desired.  To see this, let
$(q^+_{e+n},p^+_{e+n})$ be the result of applying $n$ leapfrog steps
(backward ones if $n<0$) starting at $(q^+_e,p^+_e)$. 
The probability density that $(q^+_e,p^+_e)$ will
result from mapping from a sequence to a single pair can then be written 
as follows,
considering all sequences that can contain $(q^+_e,p^+_e)$ and their
probabilities:
\beq
    \sum_{k=e-W+1}^e 
     \left[ {1 \over W}\, {\textstyle \sum_{i=k}^{k+W-1} P(q^+_i,p^+_i)} \right]
     { P(q^+_e,p^+_e) 
        \over \textstyle \sum_{i=k}^{k+W-1\rule{0pt}{8pt}} P(q^+_i,p^+_i) }
    & = & P(q^+_e,p^+_e) 
\eeq
The entire procedure therefore leaves the correct distribution invariant.

When $W>1$, the potential problem with ergodicity discussed at the end
of Section~\ref{ss:HMC} does not arise, since there is a non-zero
probability of moving to a state only one leapfrog step away, where
$q$ may differ arbitrarily from its value at the current state.

It might appear that the windowed HMC procedure requires saving all
$2W$ states in the accept and reject windows, since any one of these
states might become the new state when a state is selected from either
the accept window or reject window.  Actually, however, at most three
states need to be saved --- the start state, so that forward
simulation can be resumed after the initial backward simulation, plus
one state from the reject window and one state from the accept window,
one of which will become the new state after one of these windows is
chosen.  As states in each window are produced in sequence, a decision
is made whether the state just produced should replace the state
presently saved for that window.  Suppose the sum of the probability
densities of states seen so far is $s_i = p_1+\cdots+p_i$.  If the
state just produced has probability density $p_{i+1}$, it replaces the
previous state saved from this window with probability
$p_{i+1}/(s_i+p_{i+1})$.

I showed \citep{neal:1994} that, compared to standard HMC, using
windows improves the performance of HMC by a factor of two or more, on
multivariate Gaussian distributions in which the standard deviation in
some directions is much larger than in other directions.  This is
because the acceptance probability in equation~\eqref{e:winacc} uses
an average of probability densities over states in a window, smoothing
out the oscillations in $H$ from inexact simulation of the trajectory.
Empirically, the advantage of the windowed method was found to
increase with dimensionality.  For high-dimensional distributions, the
acceptance probability when using the optimal stepsize was
approximately 0.85, larger than the theoretical value of $0.65$ for
HMC (see Section~\ref{ss:scaling}).

These results for multivariate Gaussian distributions were obtained
with a window size, $W$, much less than the trajectory length, $L$.
For less regular distributions, it may be advantageous to use a much
larger window.  When $W=L/2$, the acceptance test determines whether
the new state is from the first half of the trajectory (which includes
the current state) or the second half; the new state is then chosen
from one half or the other with probabilities proportional to the
probability densities of the states in that half.  This choice of $W$
guards against the last few states of the trajectory having low
probability density (high $H$), as might happen if the trajectory had by then
entered a region where the stepsize used was too big.

The windowed variant of HMC may make other variants of HMC more
attractive.  One such variant (Section~\ref{ss:split}) splits the
Hamiltonian into many terms corresponding to subsets of the data,
which tends to make errors in $H$ higher (while saving computation).
Errors in $H$ have less effect when averaged over windows.  As
discussed in Section~\ref{ss:partial}, very low rejection rates are
desirable when using partial momentum refreshment.  It is easier to
obtain a low rejection probability using windows (ie, a less drastic
reduction in $\epsilon$ is needed), which makes partial momentum
refreshment more attractive.

\cite{qin-liu:2001} introduced a variant on windowed HMC.  In their
version, $L$ leapfrog steps are done from the start state, with the
accept window consisting of the states after the last $W$ of these
steps.  A state from the accept window is then selected with
probabilities proportional to their probability densities.  If the
state selected is $k$ states before the end, $k$ backwards leapfrog
steps are done from the start state, and the states found by these
steps along with those up to $W-k-1$ steps forward of the start state
form the reject window.  The state selected from the accept window
then becomes the next state with probability given by the analogue of
equation~\eqref{e:winacc}; otherwise the state remains the same.

Qin and Liu's procedure is quite similar to the original windowed HMC
procedure.  One disadvantage of Qin and Liu's procedure is that the
state is unchanged when the accept window is rejected, whereas in the
original procedure a state is selected from the reject window (which
might be the current state, but often will not be).  The only other
difference is that the number of steps from the current state to an
accepted state ranges from $L-W+1$ to $L$ (average $L-(W+1)/2$) with
Qin and Liu's procedure, versus from $L-2W+2$ to $L$ (average $L-W+1$)
for the original windowed HMC procedure, while the number of leapfrog
steps computed varies from $L$ to $L+W-1$ with Qin and Liu's
procedure, and is fixed at $L$ with the original procedure.  These
differences are slight if $W \ll L$.  Qin and Lin claim that their
procedure performs better than the original on high-dimensional
multivariate Gaussian distributions, but their experiments are
flawed.\!\footnote{In their first comparison, their method computes an
average of 55 leapfrog steps per iteration, but the original only
computes 50 steps, a difference in computation time which if properly
accounted for negates the slight advantage they see for their
procedure.  Their second comparison has a similar problem, and it is
also clear from an examination of the results (in their Table~I) that 
the sampling errors in their comparison are too large for any meaningful
conclusions to be drawn.}

\cite{qin-liu:2001} \hspace{-2pt}also introduce the more useful idea of weighting
the states in the accept\linebreak and reject windows non-uniformly, which can
be incorporated into the original procedure as well.  When mapping
from the current state to a sequence of $W$ weighted states, the
position of the current state is chosen with probabilities equal to
the weights, and when computing the acceptance probability or choosing
a state from the accept or reject window, the probability densities of
states are multiplied by their weights.  Qin and Liu use weights that
favour states more distant from the current state, which could be
useful by usually causing movement to a distant point, while allowing
choice of a nearer point if the distant points have low probability
density.  Alternatively, if one sees a window as a way of smoothing
the errors in $H$, symmetrical weights that implement a better ``low
pass filter'' would make sense.

\subsection{Using approximations to compute the trajectory}

The validity of Hamiltonian Monte Carlo does not depend on using the
correct Hamiltonian when simulating the trajectory.  We can instead
use some approximate Hamiltonian, as long as we simulate the dynamics
based on it by a method that is reversible and volume preserving.
However, the exact Hamiltonian must be used when computing the
probability of accepting the end-point of the trajectory.  There is no
need to look for an approximation to the kinetic energy, when it is of
a simple form such as \eqref{e:indK}, but the potential energy is
often much more complex and costly to compute --- for instance, it may
involve the sum of log likelihoods based on many data points, if the
data cannot be summarized by a simple sufficient statistic.  When
using trajectories of many leapfrog steps, we can therefore save much
computation time if a fast and accurate approximation to the potential
energy is available, while still obtaining exact results (apart from
the usual sampling variation inherent in MCMC).

Many ways of approximating the potential energy might be useful.  For
example, if its evaluation requires iterative numerical methods, fewer
iterations might be done than are necessary to get a result accurate
to machine precision.  In a Bayesian statistical application, a less
costly approximation to the unnormalized posterior density (whose log
gives the potential energy) may be obtainable by simply looking at
only a subset of the data.  This may not be a good strategy in
general, but I have found it useful for Gaussian process models
\citep{neal:1998,rasmussen-williams:2006}, for which computation time
scales as the cube of the number of data points, so that even a small
reduction in the number of points produces a useful speedup.

\cite{rasmussen:2003} has proposed approximating the potential energy
by modeling it as a Gaussian process, that is inferred from values of
the potential energy at positions selected during an initial
exploratory phase.  This method assumes only a degree of smoothness of
the potential energy function, and so could be widely applied.  It is
limited, however, by the cost of Gaussian process inference, and so is
most useful for problems of moderate dimensionality for which exact
evaluation of the potential energy is very costly.

An interesting possibility, to my knowledge not yet explored, would be
to express the exact potential energy as the sum of an approximate
potential energy and the error in this approximation, and to then
apply one of the splitting techniques described in
Section~\ref{ss:split} --- exploiting either the approximation's
analytic tractability (eg, for a Gaussian approximation, with
quadratic potential energy), or its low computational cost, so that
its dynamics can be accurately simulated at little cost using many
small steps.  This would reduce the number of evaluations of the
gradient of the exact potential energy if the variation in the
potential energy removed by the approximation term permits a large
stepsize for the error term.

\subsection{Short-cut trajectories --- adapting the stepsize without 
  adaptation}\label{ss:short}

One significant disadvantage of Hamiltonian Monte Carlo is that, as
discussed in Section~\ref{ss:tuning}, its performance depends
critically on the settings of its tuning parameters --- which consist
of at least the leapfrog stepsize, $\epsilon$, and number of leapfrog
steps, $L$, with variations such as windowed HMC having additional
tuning parameters as well.  The optimal choice of trajectory length
($\epsilon L$) depends on the global extent of the distribution, so
finding a good trajectory length likely requires examining a
substantial number of HMC updates.  In contrast, just a few leapfrog
steps can reveal whether some choice of stepsize is good or bad, which
leads to the possibility of trying to set the stepsize ``adaptively''
during an HMC run.

Recent work on adaptive MCMC methods is reviewed by
\cite{andrieu-thoms:2008}.  As they explain, naively choosing a
stepsize for each HMC update based on results of previous updates ---
eg, reducing the stepsize by 20\% if the previous 10 trajectories were
all rejected, and increasing it by 20\% if less than two of the 10
previous trajectories were rejected --- undermines proofs of
correctness (in particular, the process is no longer a Markov chain),
and will in general produce points from the wrong distribution.
However, correct results can be obtained if the degree of adaptation
declines over time.  Adaptive methods of this sort could be used for
HMC, in much the same way as for any other tunable MCMC method.

An alternative approach \citep{Neal:2005,Neal:2007} is to perform MCMC
updates with various values of the tuning parameters, set according to
a schedule that is predetermined or chosen randomly without reference
to the realized states, so that the usual proofs of MCMC convergence
and error analysis apply, but to do this using MCMC updates that have
been tweaked so that they require little computation time when the
tuning parameters are not appropriate for the distribution.  Most of
the computation time will then be devoted to updates with appropriate
values for the tuning parameters.  Effectively, the tuning parameters
are set adaptively from a computational point of view, but not from a
mathematical point of view.

For example, trajectories that are simulated with a stepsize that is
much too large can be rejected after only a few leapfrog steps, by
rejecting whenever the change (either way) in the Hamiltonian due to a
single step (or a short series of steps) is greater than some
threshold --- ie, we reject if $|H(q(t+\epsilon),\,p(t+\epsilon))\ -\
H(q(t),p(t))|$ is greater than the threshold.  If this happens early
in the trajectory, little computation time will have been wasted on
this unsuitable stepsize.  Such early termination of trajectories is
valid, since any MCMC update that satisfies detailed balance will
still satisfy detailed balance if it is modified to eliminate
transitions either way between certain pairs of states.

With this simple modification, we can randomly choose stepsizes from
some distribution without wasting much time on those stepsizes that
turn out to be much too large.  However, if we set the threshold small
enough to reject when the stepsize is only a bit too large, we may
terminate trajectories that would have been accepted, perhaps after a
substantial amount of computation has already been done.  Trying to
terminate trajectories early when the stepsize is smaller than optimal
carries a similar risk.

A less drastic alternative to terminating trajectories when the
stepsize seems inappropriate is to instead \textit{reverse} the
trajectory.  Suppose we perform leapfrog steps in groups of $k$ steps.
Based on the changes in $H$ over these $k$ steps, we can test whether
the stepsize is inappropriate --- eg, the group may fail the test if
the standard deviation of $H$ over the $k+1$ states is greater than
some upper threshold or less than some lower threshold (any criterion
that would yield the same decision for the reversed sequence is
valid).  When a group of $k$ leapfrog steps fails this test, the
trajectory stays at the state where this group started, rather than
moving $k$ steps forward, and the momentum variables are negated.  The
trajectory will now exactly retrace states previously computed (and
which therefore needn't be recomputed), until the initial state is
reached, at which point new states will again be computed.  If another
group of $k$ steps fails the test, the trajectory will again reverse,
after which the whole remainder of the trajectory will traverse
states already computed, allowing its endpoint to be found immediately
without further computation.

This scheme behaves the same as standard HMC if no group of $k$
leapfrog steps fails the test.  If there are two failures early in the
trajectory, little computation time will have been wasted on this
(most likely) inappropriate stepsize.  Between these extremes, it is
possible that one or two reversals will occur, but not early in the
trajectory; the endpoint of the trajectory will then usually not be
close to the initial state, so the non-negligible computation
performed will not be wasted (as it would be if the trajectory had
been terminated).

Such short-cut schemes can be effective at finding good values for a
small number of tuning parameters, for which good values will be
picked reasonably often when drawing them randomly.  It will not be
feasible for setting a large number of tuning parameters, such as the
entries in the ``mass matrix'' of equation~\eqref{e:quadK} when
dimensionality is high, since even if two reversals happen early on,
the cost of using inappropriate values of the tuning parameters will
dominate when appropriate values are chosen only very rarely.

\subsection{Tempering during a trajectory}\label{ss:temper}

Standard Hamiltonian Monte
Carlo and the variations described so far have as much difficulty
moving between modes that are separated by regions of low probability
as other local MCMC methods, such as random-walk Metropolis or Gibbs
sampling.  Several general schemes have been devised for solving
problems with such isolated modes that involve sampling from a series
of distributions that are more diffuse than the distribution of
interest.  Such schemes include parallel tempering
(\citeauthor{geyer:1991}, \citeyear{geyer:1991};
\citeauthor{earl-deem:2005}, \citeyear{earl-deem:2005}), simulated
tempering \citep{marinari-parisi:1992}, tempered transitions
\citep{neal-tt:1996}, and annealed importance sampling
\citep{neal:2001}.  Most commonly, these distributions are obtained by
varying a ``temperature'' parameter, $T$, as in
equation~\eqref{e:canT}, with $T=1$ corresponding to the distribution
of interest, and larger values of $T$ giving more diffuse
distributions.  Any of these ``tempering'' methods could be used in
conjunction with HMC.  However, tempering-like behaviour can also be
incorporated directly into trajectory used to propose a new state in
the HMC procedure.

In the simplest version of such a ``tempered trajectory'' scheme
\citep[Section 6]{neal-tt:1996}, each leapfrog step in the first half of
the trajectory is combined with multiplication of the momentum
variables by some factor $\alpha$ slightly greater than one, and each
leapfrog step in the second half of the trajectory is combined with
division of the momentum by the same factor $\alpha$.  These
multiplications and divisions can be done in various ways, as long as
the result is reversible, and the divisions are paired exactly with
multiplications.  The most symmetrical scheme is to multiply the
momentum by $\sqrt{\alpha}$ before the first half-step for momentum
(equation~\eqref{e:leap1}) and after the second half-step for momentum
(equation~\eqref{e:leap3}), for leapfrog steps in the first half of
the trajectory, and correspondingly, to divide the momentum by
$\sqrt{\alpha}$ before the first and after the second half-steps for
momentum in the second half of the trajectory.  (If the trajectory has
an odd number of leapfrog steps, for the middle leapfrog step of the
trajectory, the momentum is multiplied by $\sqrt{\alpha}$ before the
first half-step for momentum, and divided by $\sqrt{\alpha}$ after the
second half-step for momentum.)  Note that most of the multiplications
and divisions by $\sqrt{\alpha}$ are preceded or followed by another
such, and so can be combined into a single multiplication or division
by $\alpha$.
 

It is easy to see that the determinant of the Jacobian matrix for such
a tempered trajectory is one, just as for standard HMC, so its
endpoint can be used as a proposal without any need to include a
Jacobian factor in the acceptance probability.

Multiplying the momentum by an $\alpha$ that is slightly greater than
one increases the value of $H(q,p)$ slightly.  If $H$ initially had a
value typical of the canonical distribution at $T=1$, after this
multiplication, $H$ will be typical of a value of $T$ that is slightly
higher.\!\footnote{This assumes that the typical value of $H$ is a
continuous function of $T$, which may not be true for systems that
have a ``phase transition''.  Where there is a discontinuity (in
practice, a near discontinuity) in the expected value of $H$ as a
function of $T$, making small changes to $H$, as here, may be better
than making small changes to $T$ (which may imply big changes in the
distribution).}  Initially, the change in $H(q,p) = K(p) + U(q)$ is
due entirely to a change in $K(p)$ as $p$ is made bigger, but
subsequent dynamical steps will tend to distribute the increase in $H$
between $K$ and $U$, producing a more diffuse distribution for $q$
than is seen when $T=1$.  After many such multiplications of $p$ by
$\alpha$, values for $q$ can be visited that are very unlikely in the
distribution at $T=1$, allowing movement between modes that are
separated by low-probability regions.  The divisions by
$\alpha$ in the second half of the trajectory result in $H$
returning to values that are typical for $T=1$, but perhaps now in a
different mode.

If $\alpha$ is too large, the probability of accepting the endpoint of
a tempered trajectory will be small, since $H$ at the endpoint will
likely be much larger than $H$ at the initial state.  To see this,
consider a trajectory consisting of only one leapfrog step.  If
$\epsilon=0$, so that this step does nothing, the multiplication by
$\sqrt{\alpha}$ before the first half step for momentum would be
exactly cancelled by the division by $\sqrt{\alpha}$ after the second
half step for momentum, so $H$ would be unchanged, and the trajectory
would be accepted.  Since we want something to happen, however, we
will use a non-zero $\epsilon$, which will on average result in the
kinetic energy decreasing when the leapfrog step is done, as the
increase in $H$ from the multiplication by $\sqrt{\alpha}$ is
redistributed from $K$ alone to both $K$ and $U$.  The division of $p$
by $\sqrt{\alpha}$ will now not cancel the multiplication by
$\sqrt{\alpha}$ --- instead, on average, it will reduce $H$ by less
than the earlier increase.  This tendency for $H$ to be larger at the
endpoint than at the initial state can be lessened by increasing the
number of leapfrog steps, say by a factor of $R$, while reducing
$\alpha$ to $\alpha^{1/R}$, which (roughly) maintains the effective
temperature reached at the midpoint of the trajectory.


\begin{figure}[t]
\begin{center}
  \vspace*{-9pt}
  \includegraphics[width=5.5in]{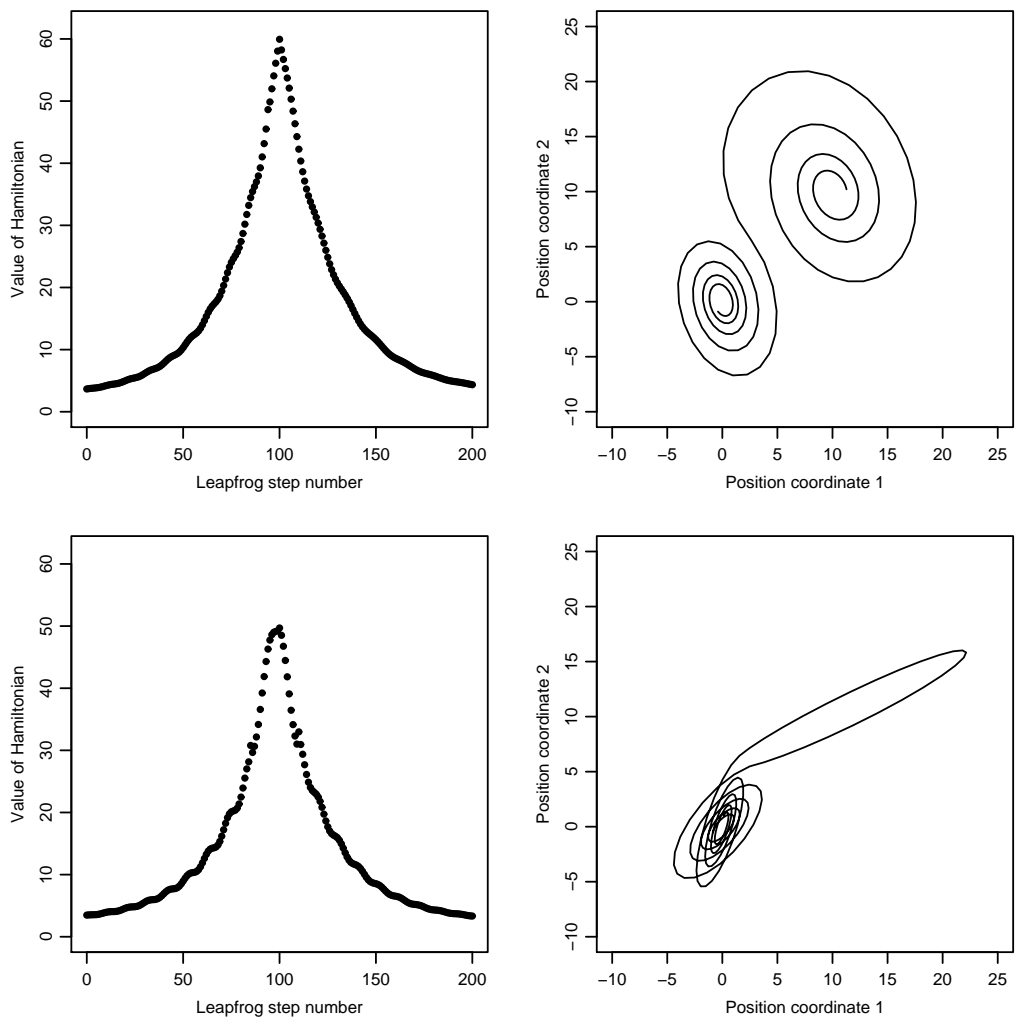}
  \vspace*{-20pt}
\end{center}
\caption{Illustration of tempered trajectories on a mixture of two Gaussians.
The trajectory shown in the top plots moves between modes;
the one shown in the bottom plots ends in the same mode.}\label{fig:temper}

\end{figure}

Figure~\ref{fig:temper} illustrates tempered trajectories used to
sample from an equal mixture of two bivariate Gaussian distributions,
with means of $[0\ 0]$ and $[10\ 10]$, and covariances of $I$ and $2I$.
Each trajectory consists of 200 leapfrog steps, done with $\epsilon=0.3$,
with tempering done as described above with $\alpha=1.04$.  The left
plots show how $H$ varies along the trajectories; the right
plots show the position coordinates for the trajectories.  The top
plots are for a trajectory starting at $q=[-0.4\ -0.9]$ and $p=[0.7\ -0.9]$,
which has an endpoint in the other mode around $[10\ 10]$.  The bottom
plots are for a trajectory starting at $q=[0.1\ 1.0]$ and $p=[0.5\ 0.8]$,
which ends in the same mode it begins in.  The change in $H$ for the
top trajectory is $0.69$, so it would be accepted with probability 
$\exp(-0.69) = 0.50$.  The change in $H$ for the bottom trajectory is
$-0.15$, so it would be accepted with probability one.

By using such tempered trajectories, HMC is able to sample these two
well-separated modes --- 11\% of the trajectories move to the other
mode and are accepted --- whereas standard HMC does very poorly, being
trapped for a very long time in one of the modes.  The parameters for
the tempered trajectories in Figure~\ref{fig:temper} where chosen to
produce easily interpreted pictures, and are not optimal.  More
efficient sampling is obtained with a much smaller number of leapfrog
steps, larger stepsize, and larger $\alpha$ --- eg, $L=20$,
$\epsilon=0.6$, and $\alpha=1.5$ gives a 6\% probability of moving
between modes.

A fundamental limitation of the tempering method described above is
that (as for standard HMC) the endpoint of the tempered trajectory is
unlikely to be accepted if the value for $H$ there is much higher that
for the initial state.  This corresponds to the probability density at
the endpoint being much lower than at the current state.
Consequently, the method will not move well between two modes with
equal total probability if one mode is high and narrow and the other
low and broad, especially when the dimensionality is high.  (Since
acceptance is based on the joint density for $q$ and $p$, there is
some slack for moving to a point where the density for $q$ alone is
different, but not enough to eliminate this problem.)  I have proposed
\citep{neal:1999} a modification that addresses this, in which the
point moved to can come from anywhere along the tempered trajectory,
not just the endpoint.  Such a point must be selected based both on
its value for $H$ and the accumulated Jacobian factor for that point,
which is easily calculated, since the determinant of the Jacobian
matrix for a multiplication of $p$ by $\alpha$ is simply $\alpha^d$,
where $d$ is the dimensionality.  This modified tempering procedure
can not only move between modes of differing width, but also move back
and forth between the tails and the central area of a heavy-tailed
distribution.

More details on these variations on Hamiltonian Monte Carlo can be
found in the R implementations available from my web page, at
\texttt{www.cs.utoronto.ca/$\sim$radford}.

\section*{Acknowledgments}

This work was supported by the Natural Sciences and Engineering Research
Council of Canada.  The author holds a Canada Research Chair in Statistics
and Machine Learning.

\bibliographystyle{apalike} 
\bibliography{ham_ref}

\end{document}